\documentclass[useAMS,usenatbib]{mn2e}

\usepackage{graphicx}
\usepackage{verbatim}
\usepackage{bm}
\usepackage{mathptmx}

\title[Single pulses of \psr]{Timing, polarimetry and physics of the bright, nearby millisecond pulsar \psr\ - a single-pulse perspective} 

\author[S. Os{\l}owski et al.]{S. Os{\l}owski$^{1,2,3,4}$\thanks{E-mail:soslowski@astro.swin.edu.au}, W. van Straten$^{1}$, M. Bailes$^{1}$, A. Jameson$^{1}$ and G. Hobbs$^{2}$\\
$^{1}$Centre for Astrophysics and Supercomputing, Swinburne University of Technology, Mail H39, PO Box 218, VIC 3122, Australia\\
$^{2}$CSIRO Astronomy and Space Sciences, Australia Telescope National Facility, PO Box 76, Epping, NSW 1710, Australia\\
$^{3}$Max-Planck-Institut f{\"u}r Radioastronomie, Auf dem H{\"u}gel 69, D-53121 Bonn, Germany \\
$^{4}$Fakult{\"a}t f{\"u}r Physik, Universit{\"a}t Bielefeld, Postfach 100131, D-33501 Bielefeld, Germany \\}

\begin{document}

\newcommand{\msun}{\mbox{M$_{\odot}$}}
\newcommand{\rsun}{\mbox{R$_{\odot}$}}
\newcommand{\aopx}{\mbox{$\Delta_{\pi \rm M}$}}
\newcommand{\shap}{\mbox{$\Delta_{\rm S}$}}
\newcommand{\psr}{\mbox{PSR J0437$-$4715}}

\newcommand{\pb}{\mbox{$P_{\rm b}$}}
\newcommand{\pbdot}{\mbox{$\dot{\pb}$}}

\newcommand{\wvect[1]}{{\bsf #1}}
\newcommand{\bvect[1]}{{\bsf #1}$_{\bm 0}$}

\let\oldhat\hat
\renewcommand{\vec}[1]{\mathbf{#1}}
\renewcommand{\hat}[1]{\oldhat{\mathbf{#1}}}

\date{Accepted . Received ; in original form}

\pagerange{\pageref{firstpage}--\pageref{lastpage}} 
\pubyear{2011}

\maketitle

\label{firstpage}
  
\begin{abstract}
	Single pulses from radio pulsars contain a wealth of information about emission and propagation in the magnetosphere and insight into their timing properties. It was recently demonstrated that single-pulse emission is responsible for limiting the timing stability of the brightest of millisecond pulsars. We report on an analysis of more than a million single-pulses from \psr\ and present various statistical properties such as the signal-to-noise ratio ($S/N$) distribution, timing and polarimetry of average profiles integrated from subpulses with chosen $S/N$ cut-offs, modulation properties of the emission, phase-resolved statistics of the $S/N$, and two dimensional spherical histograms of the polarization vector orientation. The last of these indicates the presence of orthogonally polarised modes (OPMs). Combined with the dependence of the polarisation fraction on the $S/N$ and polarimetry of the brightest pulses, the existence of OPMs constrains pulsar emission mechanisms and models for the plasma physics in the magnetosphere.

\end{abstract}

\begin{keywords}
pulsars: general -- pulsars: individual (\psr)

\end{keywords}

\section{Introduction} % - 1000 words}
\label{section::intro}

Radio pulsars are highly magnetized, quickly spinning neutron stars that emit radio waves. A non-thermal emission mechanism, not yet understood despite decades of study, is responsible for emission in two beams aligned with the magnetic field axis of the pulsar. As the spin axis is misaligned with the magnetic field axis, the radio beams can sweep past Earth, producing a light-house like effect of emission across the electromagnetic spectrum. Pulsars are extremely versatile laboratories of physics and observing them provides unique insights into physics at many scales. Neutron stars probe fundamental interactions via the ultra-dense matter and solid state physics  \citep[e.g.,][]{1999ApJ...512..288T,2013Sci...340..448A}. Plasma physics in highly magnetized environments is relevant in the pulsar magnetosphere \citep[e.g.,][]{2004MNRAS.353..270C}. Pulsars allow for tests of non-linear strong-field gravity, probing regions of phase space inaccessible by other means \citep[e.g.,][]{2004NewAR..48..993K,2014Natur.505..520R}. Pulsar Timing Arrays are recently gaining momentum in their quest for the detection of gravitational waves  \citep[e.g.,][]{1978SvA....22...36S,1990ApJ...361..300F,2013Sci...342..334S}. Other types of experiments possible with pulsar  observations include the derivation of a distance scale, study of the interstellar tenuous plasma and large-scale magnetic fields \citep{2006ApJ...642..868H,2008ApJ...685L..67D,2008MNRAS.386.1881N,2009ASTRA...5...43B,2012MNRAS.427..664S}, independent measurements of masses in the Solar system \citep{2010ApJ...720L.201C}, development of a pulsar-based time-scale \citep{2012MNRAS.427.2780H}, and many more \citep[e.g.,][]{2004NewAR..48.1413C}.

Many of the aforementioned experiments are based on pulsar timing methodology, i.e., estimation of the mean time of arrival (ToA) of the pulse train. The precision attainable in such experiments is dependent on the pulse profile (i.e., the longitude resolved light curve of a pulsar), spin period, brightness of the pulsar, and so called timing noise, that is any unexplained deviations of the pulse train arrival times from simple physical models\footnote{There is no widely accepted definition of timing noise and some authors include in this term only processes with certain spectral characteristics}. The most precise timing experiments are based on the observations of millisecond pulsars (MSPs), i.e., pulsars that have been spun up by mass accretion and the associated transfer of angular momentum from the companion star \citep[][]{1982Natur.300..728A} or directly through the accretion-induced collapse of a white dwarf \citep{2014MNRAS.438L..86F}. 

Another way of studying pulsars is by analysing the single pulses emitted by the pulsar.  While the average pulse profile is generally stable over time \citep{1975ApJ...198..661H}, every single pulse is different \citep{1968Sci...160..758D}. Their amplitudes can be modulated temporally with or without a drift in phase \citep{1970Natur.227..788C,2003A&A...407..273E}. Some pulsars give off  giant pulses \citep[e.g.,][]{1968Sci...162.1481S,1969Natur.221..453C,2006ApJ...640..941K,2009MNRAS.395.1972V} and micropulses \citep{2001ApJ...549L.101J}, and some have high nulling fractions, that is the fraction of time when the emission of the pulsar is not detectable \citep{1970Natur.228...42B}. The latter effect may be connected with the magnetospheric state and the spin-down rate \citep{2006Sci...312..549K,2010Sci...329..408L}. The emission is not only often highly polarized but can also exhibit two orthogonally polarised modes \citep[OPMs][]{1975ApJ...196...83M,1976Natur.263..202B,1978ApJ...223..961C,1980ApJS...42..143B,1984ApJS...55..247S}. MSPs are typically less luminous and fainter than more slowly rotating pulsars. They also spin much more quickly, reducing the signal-to-noise ratio ($S/N$) of individual pulses and increasing the volume of data by two or three orders of magnitude per unit observing time. For these reasons, single-pulse studies of MSPs are much more difficult. Such studies can shed light on the elusive nature of the pulsar emission mechanism, can help show whether there is a continuity of single pulse properties from MSPs to slower pulsars, can provide an insight into plasma physics, and can also be used to determine the fundamental limits of timing precision.

Empirical and theoretical considerations of single-pulse properties and studies of slow pulsars' single-pulse emission have been undertaken in the past \citep[e.g., most recently,][]{2001A&A...379..270K,2002A&A...391..247K,2003A&A...404..325K,2003A&A...407..655K,2003MNRAS.344L..69K,2004MNRAS.352..915M,2006MNRAS.365..638M,2007A&A...462..257B,2012MNRAS.424..843W}. The sample of MSP single-pulse studies is more limited but still numerous \citep{1996ApJ...457L..81C,2001ApJ...546..394J,2003A&A...407..273E,2004ApJ...602L..89J,2006ApJ...640..941K,2012MNRAS.423.1351B,2013MNRAS.430.2815Z}.

A perfect candidate among MSPs for single pulse studies is \psr, discovered in the 70-cm Parkes Survey \citep{1993Natur.361..613J}, the brightest MSP known; a number of such studies have been undertaken previously \citep{1993Natur.361..613J,1997ApJ...475L..33A,1998ApJ...498..365J,1998ApJ...501..823V,2000ApJ...543..979V}.  Owing to the progress of technology and methodology, we present a study of a larger data set, recorded over a wider band, with fully calibrated polarization data.

In this paper, we focus on the relation between the single-pulse properties and the timing of \psr\ as well as on the polarimetric properties of the single pulses. In Section \ref{obs} we describe our observations including the hardware used and details of data processing; section \ref{tip} focuses on an analysis of the total intensity properties of \psr; the polarimetric properties are presented in Section \ref{pip}; and we summarize our findings in Section \ref{conclusions}.

\section{Observations and data processing}
\label{obs}

Observations of \psr\ were recorded in two approximately hour-long sessions in 2011 November using the Parkes 64-m radio telescope and the central beam of the 21 cm multibeam receiver \citep{1996PASA...13..243S}. We used the CASPER Parkes Swinburne Recorder (CASPSR), an 8-bit baseband recorder and a digital signal processing system, capable of real-time phase-coherent dispersion removal over a band up to $400$~MHz wide. {\sc dspsr} \citep{2011PASA...28....1V} was used to process the real-sampled voltages from a $300$~MHz wide band\footnote{The receiver used during the observing sessions provided only $300$~MHz of bandwidth} centred at $1382$~MHz.  In total 117.3 min of data were recorded, or about 1.2 million pulsar rotations, each resolved into 1024 phase bins. A 512 channel filterbank was created while performing phase-coherent dispersion removal assuming a dispersion measure of $2.64476$ \citep{2008ApJ...679..675V}. Inter-channel dispersion delays were removed before dividing the data into single pulses and recording the data on a Redundant Array of Independent Discs. We point out that, if the ephemeris aligns the peak of the total intensity at phase zero, then the reference phase in {\sc dspsr} needs to be offset to 0.5 turns. Otherwise every single-pulse will be split between two data files. All subsequent processing was performed using the {\sc PSRchive} software suite \citep{2004PASA...21..302H,2012AR&T....9..237V}. The data were converted to PSRFITS format \citep{2004PASA...21..302H} and checked for incorrect header parameters.

Narrow-band radio frequency interference (RFI) was removed by applying a median filter, i.e., comparing the total  flux density in each channel with that of its 49 neighbouring channels. We attempted several methods of post-folding removal of impulsive RFI but all of them removed some of the brightest pulses, rendering automatic removal impractical. \citet{1998ApJ...498..365J} demonstrated that single pulses of \psr\ are much narrower than the average pulse profile. This allowed us to pre-select potential RFI by searching for wide pulses and visually inspecting all the pulses broader than 54 phase bins, the expected maximum width of a sub-pulse\footnote{By a sub-pulse we refer to a single emission event within a single pulse, that is a single pulse can consist of multiple sub-pulses}. Also, the top 1\% of pulses sorted by $S/N$ were visually scrutinized as the data with unexpectedly high S/N is likely to be RFI. This inspection found that only 0.01\% of data is affected by strong impulsive RFI  and thus we ignore its presence throughout the analysis. We note that since its incorporation into the {\sc dspsr} package, the method of spectral kurtosis \citep{2007PASP..119..805N,2010PASP..122..595N} is becoming increasingly popular for rejection of RFI during pulsar observations. Unfortunately, it cannot be applied to our data. The brightest pulses of \psr\ are difficult to distinguish from RFI using methods that excise signal above certain thresholds (e.g., total power or kurtosis) because the sub-pulse width and dispersive smearing are smaller than the time-scale over which these statistics are computed.

\begin{figure}
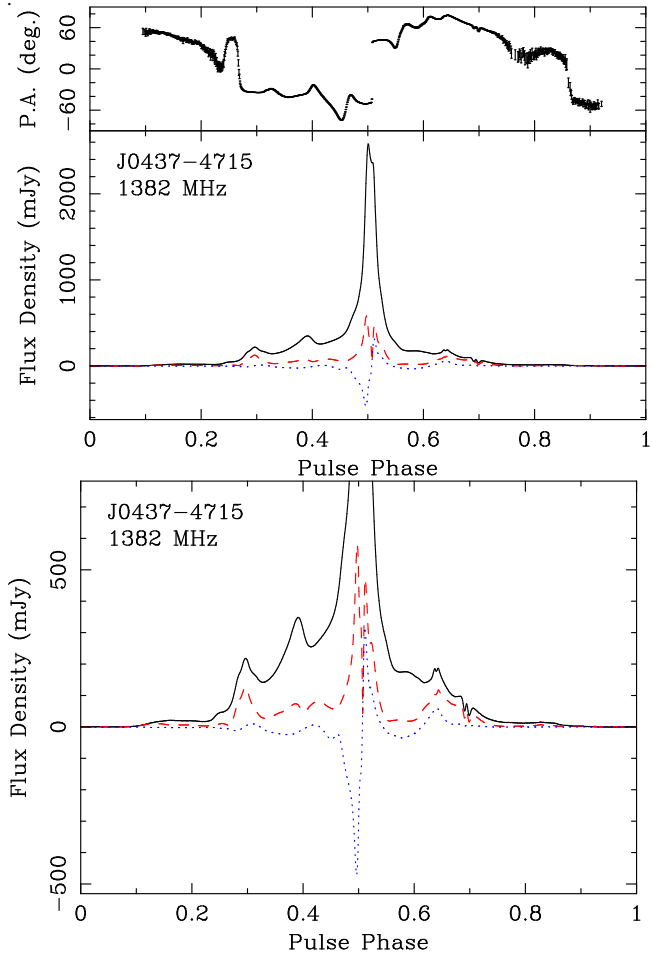
.
\includegraphics[angle=-90,width=\columnwidth]{figures/template_Stokes_caspsr_r52}
\includegraphics[angle=-90,width=\columnwidth]{figures/template_Stokes_caspsr_r52_crop}
\caption[High S/N template from CASPSR]{Top: the high S/N ($\sim20,000$) template for \psr\ created from $8.5$ hours of observations is shown along with the polarization angle in the top panel. In the bottom panel, the solid black line represents total intensity, the dashed red line corresponds to linear polarisation while the dotted blue line represents the circular polarisation. Bottom: here we show the same profile but zoomed in to facilitate examination of the details of the polarisation curves in the lower profile.  \label{Fig::std}}

\end{figure}

In order to perform polarimetric calibration using measurement equation template matching \citep[METM,][]{2013ApJS..204...13V} we applied the following procedure: an eight-hour long observing track of \psr\ observed within a month of the observing session in 2011 November was calibrated using the measurement equation modelling technique \citep{2004ApJS..152..129V}. A high S/N calibrated template, consistent with data presented by \citet{2011MNRAS.414.2087Y}, was obtained and this is presented in Fig. \ref{Fig::std}. For each hour session of our main data set, five minute integrations were formed by integrating single pulses in a hierarchical manner to minimize rounding error. These five minute integrations were used to derive properties of the receiver at the time of observation. This information was used to calibrate the single pulses. See \citet{2013ApJS..204...13V} for details of the outlined procedure. Observations of the Hydra A radio galaxy, which is assumed to have a constant  flux density of $43.1$\,Jy at $1400$\,MHz and a spectral index of $-0.91$ \citep{1968ARA&A...6..321S}, were used to calibrate the  flux density scale. 

\section{Total intensity properties}
\label{tip}

We begin the data analysis with a focus on the total intensity of single pulses. We present the distribution of the single-pulse intensity and dependence of the average pulse profile on the S/N threshold for data inclusion before discussing the impact of the pulse energy distribution on timing properties of \psr. We then analyse the implications of phase-resolved flux density distributions on plasma physics in the emitting region, and finally discuss the quasi-periodic stationary and drifting intensity modulation.

\subsection{Distribution of instantaneous signal-to-noise ratio}
\label{section::SNRdist}
The flux density distribution of \psr\ single pulses has been presented previously by \citet{1998ApJ...498..365J} and is consistent with our findings. Here we present the distribution of instantaneous signal-to-noise ratio, as measured according to two separate definitions. The first one is:
\begin{equation}
S/N=\frac{\sum^{N_{\rm on}}_{\rm i=1}\left(F_i-B\right)}{\sqrt{N_{\rm on}}\sigma_{\rm off}}\,{\rm ,}
\label{eq::s/n_trad}
\end{equation}
where $F_i$ is the pulse  flux density in the i-th on-pulse bin; $B$ is the mean off-pulse  flux density; $N_{\rm on}$ is the number of on-pulse phase bins; and $\sigma_{\rm off}$ is the off-pulse root mean square (rms)  flux density. The signal-to-noise ratio defined in this way is independent of the number of the phase bins used to resolve the pulse period. We note that the S/N of a narrow pulse with high  flux density can be the same as a broad pulse with smaller  flux density, making the definition of S/N sensitive to RFI which is often broader than the single pulses. Therefore, we checked our results using an alternative definition:
\begin{equation}
S/N_{\rm peak}=\frac{F_{\rm max}-B}{\sigma_{\rm off}}\,{\rm ,}
\label{eq:s/n}
\end{equation}
where $F_{\rm max}$ is the maximum  flux density;  $S/N_{\rm peak}$ is a phase resolution-dependent quantity but it is more useful for finding high  flux density spikes in the data. %We note that both of these definitions are based on the off-pulse noise estimates and as such, do not take the stochastic wideband impulse modulated self-noise (SWIMS, also referred to as the pulse or phase jitter in the literature) into account in the definition of $\sigma_{\rm off}$.

The pulsar signal travels through the interstellar medium on the way to Earth. This has several consequences, such as dispersion and interstellar scintillation, the latter of which causes the apparent  flux density to vary significantly, both in time and frequency \citep[see][for a review]{1990ARA&A..28..561R}. By chance, the apparent  flux density was relatively stable during observations presented in this work and variations were within a factor of two. To facilitate comparison of single pulses with similar intrinsic S/N we use the normalized quantity $S/N_n=\frac{S/N}{\left<S/N\right>}$, where $\left<S/N\right>$ is the mean $S/N$ in a minute \citep[compared to a refractive scintillation time-scale of the order of 1000 s,][]{2006A&A...453..595G} centred on the given single pulse.

\begin{figure}
\includegraphics[width=\columnwidth]{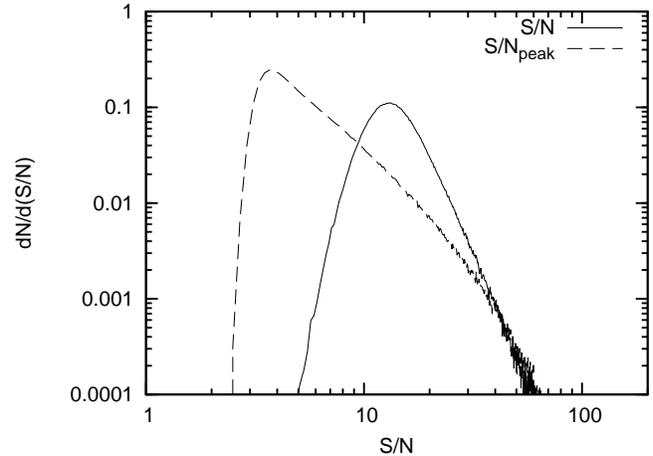}
\caption[Observed instantaneous signal to noise ratio]{Observed instantaneous signal-to-noise ratio. The solid line is the distribution of $S/N$ while the dashed line represents the distribution of $S/N_{\rm peak}$. Overlap of the two distributions at high S/N range suggests that the brightest pulses are also the most narrow. \label{Fig::snr}}
\end{figure}

\begin{figure}
\includegraphics[angle=-90,width=\columnwidth]{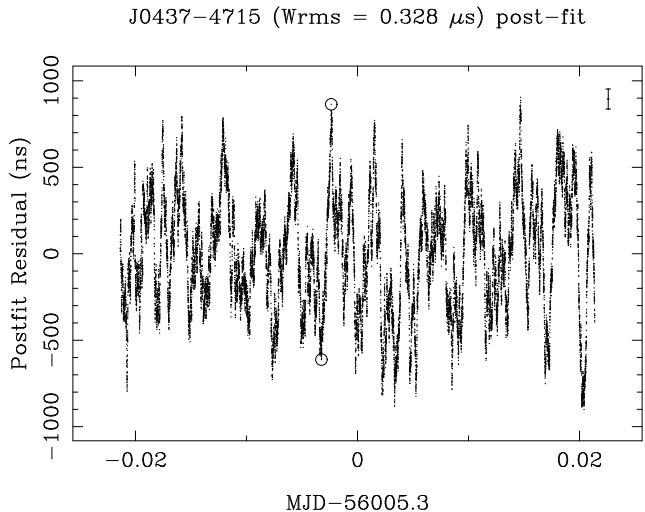}
\caption[Running timing residual estimate for one minute integrations]{Running timing residual estimate for one minute integrations. For practical plotting reasons, only every 25th point is plotted and no measurement errors are included. We also point out that the data points within one minute are not independent as overlapping date are used to derive the ToAs. The two marked points are taken $76.4\;{\rm s}$ apart and the difference in residual ToA is $1500\;{\rm ns}$ while the mean ToA estimation error is only $58$~ns. To facilitate comparison, we added a point with an error bar of this size in the top right corner of the Figure. \label{Fig::running}}
\end{figure}

The peak of $S/N_{\rm peak}$ distribution, as seen in Fig. \ref{Fig::snr}, corresponds to the expected maximum amplitude of the noise in every pulse as we have 1024 phase bins and thus expect a 3 standard deviations peak in each observation. The instantaneous median unnormalized $S/N$ is $10$. This immediately implies that stochastic wideband impulse-modulated self-noise (SWIMS) will be significant in the observations of this pulsar \citep{2011MNRAS.418.1258O}. Due to heteroscedasticity and temporal correlation of SWIMS, the ToA estimates of average pulse profiles of \psr\ are biased \citep{2010arXiv1010.3785C,2011MNRAS.418.1258O,2012MNRAS.420..361L} increasing the rms of the timing residuals (i.e. the difference between the modelled ToAs and the observations) by a factor of four. \citet{2011MNRAS.418.1258O} and \citet{2013MNRAS.430..416O} have presented a statistical method of correcting this bias by characterising the SWIMS in the pulse profiles. In the following two subsections we present an investigation of a different approach to detecting the impact of SWIMS. 

\subsection{Impact of a single pulse on the ToA estimate}

This and the next two subsections focus on total intensity properties of \psr\ which relate to timing. We begin here by measuring the impact of a single pulse on the ToA estimate before moving on to an analysis of the evolution of selectively integrated pulse profiles and their timing properties.

As demonstrated in section~\ref{section::SNRdist}, the instantaneous S/N of single pulses from \psr\ often exceeds unity. In such a case, we expect the effect of a single pulse on the ToA estimated from an integration of some length to be directly measurable. In this section we quantify the average impact of a single pulse on the ToA estimate. To achieve this goal we implemented the following procedure:
\begin{enumerate}
\item we integrated one minute of data and formed an average pulse profile,
\item this profile was timed against the template profile presented in Fig. \ref{Fig::std},
\item we then removed the first single pulse from this minute of data and added the next single pulse that follows the original minute of integration,
\item finally we go back to the first step and average the profile.
\end{enumerate}
We have analysed in this way all of the single pulses in one hour of data, thus forming a ``running ToA'' presented in Fig. \ref{Fig::running}. A typical ToA error estimate equals $58$~ns. Note that we present only every 25th resulting ToA for practical reasons. The data points in this plot are not independent as two neighbouring ToAs are derived from data with $99.76\%$ of overlap.

We highlight two points in this plot around the $-0.0025$ abscissa value. These two timing residuals are separated by only $76.4\;{\rm s}$ and yet the difference between their estimates is $1500\;{\rm ns}$, nearly 26 times the typical ToA error estimate. We can use these values to directly measure the bias introduced by SWIMS on the ToA estimate. Every single pulse between the two highlighted ToAs has affected the estimated time of arrival by $0.11\;{\rm ns}$ on average. While this number seems negligible, it implies that if an observer decided to postpone both the start and finish of a 1-minute observation by one second, the estimated ToA would change by $19.1\;{\rm ns}$ on average. With next-generation radio telescopes, such as FAST  \citep{FAST} or the Square Kilometre Array (SKA) \citep{2010SPIE.7733E..35S} an observer might be satisfied with the achieved S/N ratio after a one-minute integration; however, the effect of SWIMS may dominate ToA estimates at frequencies similar to that of the observations presented in this work.. At other frequencies the relative importance of SWIMS will likely change. 

\begin{figure}
\includegraphics[angle=-90,width=\columnwidth]{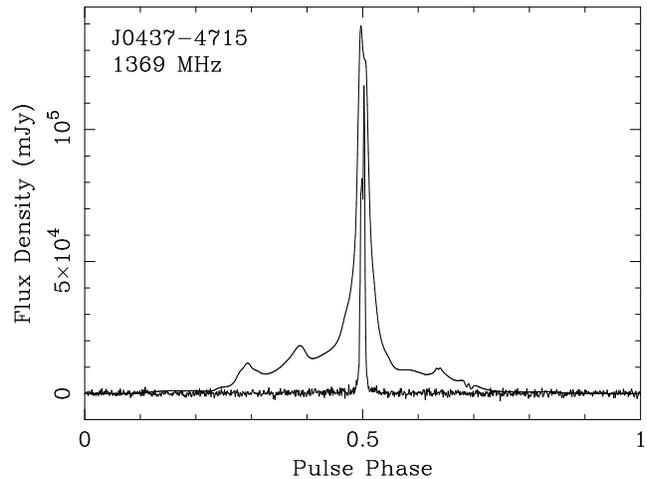}
\caption[Comparison of a single pulse with an average pulse profile]{Comparison of a single pulse with an average pulse profile. The  flux density scale matches the single pulse and the  flux density of the average pulse profile has been multiplied by a factor of 45 for clarity. Total  flux density contained in the single pulse equals $636.8$~Jy compared to the total  flux density in the average pulse profile of the observing session this pulse comes from of $155.8$~Jy. \label{Fig::single_vs_avg}}
\end{figure}

Simulations performed with the ``psrover'' application within {\sc PSRchive} show that a single pulse can affect the ToA even if it has only a fraction of the  flux density of the average profile, especially if it is a wide pulse. In case of \psr\, the subpulses are quite narrow but can have a very high  flux density, as demonstrated in Fig. \ref{Fig::single_vs_avg} and shown previously by \citet{1998ApJ...498..365J}. For the sake of visibility of the average pulse profile, the template's  flux density was artificially increased by a factor of 45. The energy in a single pulse can be larger than in the average pulse profile and concentrated over a much narrower phase range. The average pulse profile will be measurably different depending on the inclusion of this single pulse in the average, impacting the template matching algorithm used to derive the ToAs. 

\subsection{Average pulse profile dependence on sub-pulse S/N}
\label{avg_SN}

\begin{figure}
\includegraphics[angle=-90,width=\columnwidth]{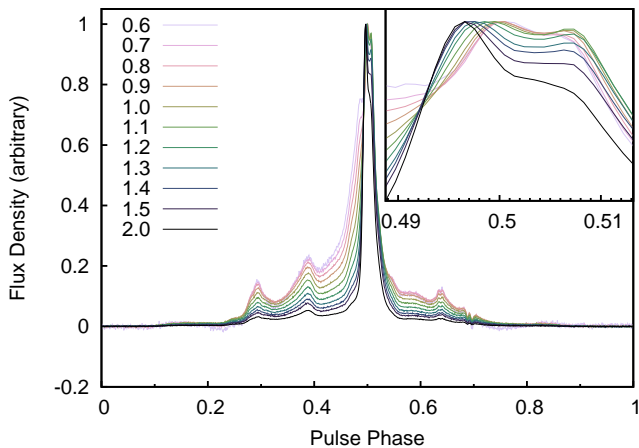}
\caption[Average pulse profile for selective integration]{Average pulse profile integrated using only data within given instantaneous normalized S/N ranges. The profiles were normalised to have the same peak  flux density. The numbers in the label denote to the top end of the normalised S/N range. We used progressively darker colours to denote the different ranges. The lower limit of any range is the upper limit of the previous range, or $0.0$ in the case of range denoted as $0.6$. The inset in top-right corner zooms in on the peak region, i.e., from phase $0.488$ to $0.516$. \label{Fig::waterfall}}
\end{figure}

We first present the dependence of the pulse profiles on the thresholds of $S/N_n$ applied to the data used to create of these profiles. Extending the ideas of \citet{1983ApJ...265..372K}\footnote{The authors referred to the process of selective integration as gating which nowadays is often used to describe selecting a phase range of the pulse profile.}, we divide our data into the following $S/N_n$ ranges: $[0.0;0.5)$; $[0.5,0.6)$; $[0.6,0.7)$; $[0.7,0.8)$; $[0.8,0.9)$; $[0.9,1.0)$; $[1.0,1.1)$; $[1.1,1.2)$; $[1.2,1.3)$; $[1.3,1.4)$; $[1.4,1.5)$; $[1.5,2.0)$; $[2.0;4.0)$; $[4.0:8.0)$; $[8.0,16.0)$; and $>16.0$. The average pulse profiles in a subset of these $S/N_n$ ranges are plotted in Fig. \ref{Fig::waterfall}. The averaged profiles were scaled to have the same peak intensity to facilitate comparison. The brightest pulses tend to occur near the peak of the template profile intensity. \citet{2012ApJ...761...64S} have seen similar behaviour for \mbox{PSR J1713+0747}, which exhibits a correlation between the $S/N$ of the single pulses and their time of arrival. 

The investigation of Fig. \ref{Fig::waterfall} reveals that the two components distinguishable in the very centre of the profile are of nearly equal amplitudes when only the weakest pulses are integrated and the peak dominant in the average profile becomes more pronounced in the strongest pulses. In addition, in the average pulse profile formed from the weakest pulses, an additional component on the leading edge becomes visible. This component, present around pulse phase 0.47, can be seen in the template profile and should be included when modelling the pulse profile with analytical functions, see the bottom panel of Fig.~\ref{Fig::std}.

\begin{figure}
\includegraphics[angle=-90,width=\columnwidth]{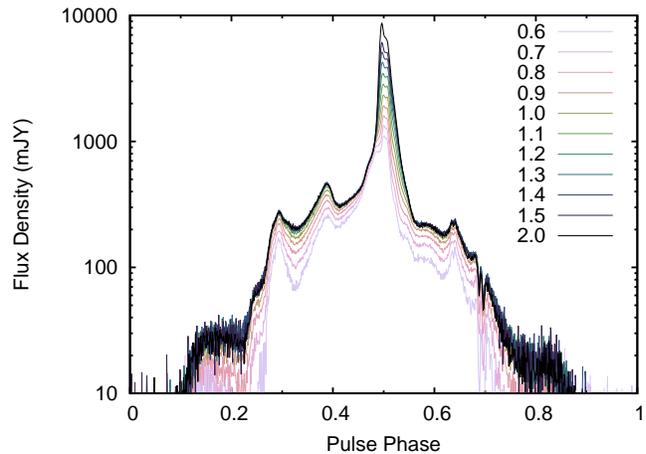}
\caption[Average pulse profile for selective integration]{Average pulse profile integrated using only data within given instantaneous normalized S/N ranges. This figure is similar to Fig. \ref{Fig::waterfall}, but this time the profiles were not re-normalised and we used logarithmic  flux density scale to bring out the details in the profile wings. \label{Fig::waterfall_wings}}
\end{figure}

We note that all of the profiles, even those constructed from the brightest pulses, exhibit the structure in the ``wings''\footnote{We use the term ``wings'' to refer to the phase region where the  flux density is much lower than at the peak of the profile, e.g., from the first minima of the pulse profile on either side of the peak below  flux density of $500$~mJy in Fig.~\ref{Fig::std}.} identical to that of the average profile (see Fig. \ref{Fig::waterfall_wings}). This structure persists in all of the S/N ranges. Visual inspection of a number of the brightest single pulses, conducted during the search for RFI described in section~\ref{obs}, reveals that not a single one of them has any visible emission in the wings phase region, and the vast majority of the single pulses are detected only in 20 phase bins near phase 0.5. The persistent presence of wings in the average pulse profile of these brightest pulses implies the presence of low-level emission, which is revealed only by averaging away the radiometer noise. This suggests that the pulsar magnetosphere is local, i.e., the presence of a bright single pulse does not affect other regions of magnetosphere, which continue to emit without being affected by any physical process that generated the bright emission. 

\subsection{Timing properties as a function of sub-pulse S/N}

Because SWIMS begins to impact the ToA when the instantaneous S/N exceeds unity, it may be possible to limit the amount of such noise in the data by rejecting the brightest single pulses as identified by their normalized S/N. While such a procedure will reduce the S/N of the average pulse profiles, it may be beneficial as the noise will be closer to having equal variance in each pulse phase bin and being temporally uncorrelated; that is, the presence of SWIMS will be limited. We note that this will not completely remove the spectral correlations in the noise. In Table \ref{t::rms}, we present the rms of the timing residuals for average profiles formed by integrating for one minute but including only single pulses with normalised S/N in the following ranges: $S/N_n<0.5$; $S/N_n<1.0$; $S/N_n<2.0$; $S/N_n<4.0$; $S/N_n<8.0$ and the complementing ranges: $>0.5$; $>1.0$; $>2.0$; $>4.0$; $>8.0$; In this case we have taken the identification of RFI into account and used only the data that was not flagged as affected by RFI. The table also contains the fraction of pulses that have been classified as belonging in the respective range.

\begin{table}
\caption[Rms of the timing residual for selective integration]{Rms of the timing residual for data formed by excluding single pulses in given $S/N_n$ ranges. The rms of the timing residual is given in nanoseconds. The row highlighted in bold corresponds to standard timing technique.}
\begin{tabular}{ c | c  c c }
	Included range	&	rms timing residual	&	$\chi^2/{\rm d.o.f.}$ & Fraction of pulses \\
		\hline
$S/N_n < 0.5$ & $4998$	 & $22.3$ & 0.0054 \\
$S/N_n < 1.0$ & $449$	 & $14.3$ & 0.6011 \\
$S/N_n < 2.0$ & $331$	 & $31.1$ & 0.9476 \\
$S/N_n < 4.0$ & $314$	 & $34.7$ & 0.9673\\
$S/N < 8.0$ & $312$	 & $34.5$ & 0.9679\\
$\bm{S/N_n > 0.0}$ & $\bm{338}$       & $\bm{39}$ & $\bm{1.000}$ \\
$S/N_n > 0.5$ & $316$	 & $34.4$ & 0.9625 \\
$S/N_n > 1.0$ & $340$	 & $44.9$ & 0.3669 \\
$S/N_n > 2.0$ & $931$	 & $181.3$ & 0.0204 \\
$S/N_n > 4.0$ & $3790$	 & $157$ & 0.0007 \\
\end{tabular}
\label{t::rms}
\end{table}

The rms of the timing residual is marginally reduced for the data formed by rejecting the single pulses of $S/N_n$ greater than 2, 4 or 8. The  $\chi^2/{\rm d.o.f.}$ of the timing model is lower in these cases as well as for any other case when an upper limit on $S/N_n$ is imposed. As the high instantaneous $S/N$ pulses are removed from data, the relative contribution of SWIMS is reduced along with a reduction of the average $S/N$. This causes the ToA measurement uncertainty estimates to be both larger and more realistic, thus reducing the $\chi^2/{\rm d.o.f.}$. 
On the opposite end of the spectrum, the brighter the included single pulses are, the ToA estimates become dominated by SWIMS and the timing is strongly affected, visible in both the rms of the timing residual and the $\chi^2/{\rm d.o.f.}$ of the fit. While it is not possible to significantly improve the rms of the timing residual by selectively rejecting single pulses in this data set, it may be possible for other pulsars with a different distribution of single-pulse flux densities.

\subsection{Observed electric field intensities and plasma physics}

After decades of research and despite many viable possibilities presented in the literature, the pulsar emission mechanism remains elusive. Following \citet{2003MNRAS.343..512C} we list the proposed linear processes including, but not limited to, linear acceleration and maser curvature emission \citep{1995MNRAS.276..372L,1996ASPC..105..139M,2003PPCF...45..523M}, relativistic plasma emission \citep{1996ASPC..105..139M,1996ASPC..105..147A}, and a streaming instability into an escaping mode \citep{2002MNRAS.337..422G}. Non-linear processes have been proposed by several groups.  The possibilities include direct conversion of plasma turbulence into electromagnetic emission \citep{1997ApJ...483..402W,1998ApJ...506..341W}, soliton collapse \citep[][and references therein]{2006MNRAS.369.1469A} and an antenna mechanism \citep{1992JGR....9712029P,2000ralw.conf...27C}. Most models of emission depend on the presence of dense plasma in the pulsar magnetosphere, exceeding the value expected in the classical model of \citet{1969ApJ...157..869G} by orders of magnitude. A possible solution is that additional plasma is created in pair cascades in the acceleration gaps: polar \citep{1971ApJ...164..529S,1975ApJ...196...51R}, outer \citep{1986ApJ...300..500C,1996ApJ...470..469R} or slot \citep{2003ApJ...588..430M}. Observation of highly energetic emission from pulsars suggest that the polar gap geometry is only viable for older pulsars while more energetic pulsars create plasma in either outer or slot gaps \citep{2010ApJ...716L..85R} with the first one being favoured in a different study \citep{2010ApJ...714..810R}.

An important aspect of emission is the interaction between plasma waves, driving particles and the background plasma in the magnetosphere. As pointed out in a series of papers by Cairns and collaborators \citep{2001ApJ...563L..65C,2003MNRAS.343..523C,2003MNRAS.343..512C}, field statistics provide insight into plasma behaviour in the pulsar's magnetosphere and these authors discuss two competing theories of such interactions.  In brief, self-organized criticality \citep[SOC;][]{1988PhRvA..38..364B,1996hnw..book.....B} describes systems interacting self-consistently without any preferred distance- or time-scales and predicts power-law distributions of intensities with indices typically in the range $0.5$ to $2.0$. Stochastic growth theory \citep[SGT;][]{1992SoPh..139..147R,1995PhPl....2.1466R,1993ApJ...407..790R,2001JGR...10629515C} in turn describes self-consistently interacting systems where the interactions take place in an independent homogeneous medium and introduce  distance- and time-scales. This theory predicts a log-normal distribution of electric field.

The field statistics have been studied so far for the Vela pulsar in the aforementioned series of papers by Cairns and collaborators, as well as for \mbox{PSR J1644$-$4559} and \mbox{PSR J0953+0755} in a later work \citep{2004MNRAS.353..270C}. \citet{2006ARep...50..915S} demonstrated that the intensity of \mbox{PSR J0953+0755} is distributed log-normally at certain phases. The Crab pulsar is known to emit giant pulses which are very narrow and highly energetic emission events \citep{1968Sci...162.1481S,2003Natur.422..141H}. To check if MSPs also show such pulses, \citet{2006ApJ...640..941K} studied four MSPs. The authors found that \mbox{PSR J0218+4232} emits giants pulses with a power-law distribution of energies. \citet{2012MNRAS.423.1351B} studied the field statistics for a sample of 315 pulsars, including two MSPs, \mbox{PSR J1439$-$5501} and \mbox{PSR J1744$-$1134}, but these distributions were not phase-resolved. The MSPs were found to have non-Gaussian and non-log-normal distribution of energies.  Another related study of MSPs was done previously by \citet{1996ApJ...457L..81C} who studied the giant pulses from \mbox{PSR J1939+2134}. They found that the amplitudes of these pulses follow power-law distributions for the pulses from both the main- and inter-pulse. 

\begin{figure}
\includegraphics[angle=0,width=\columnwidth]{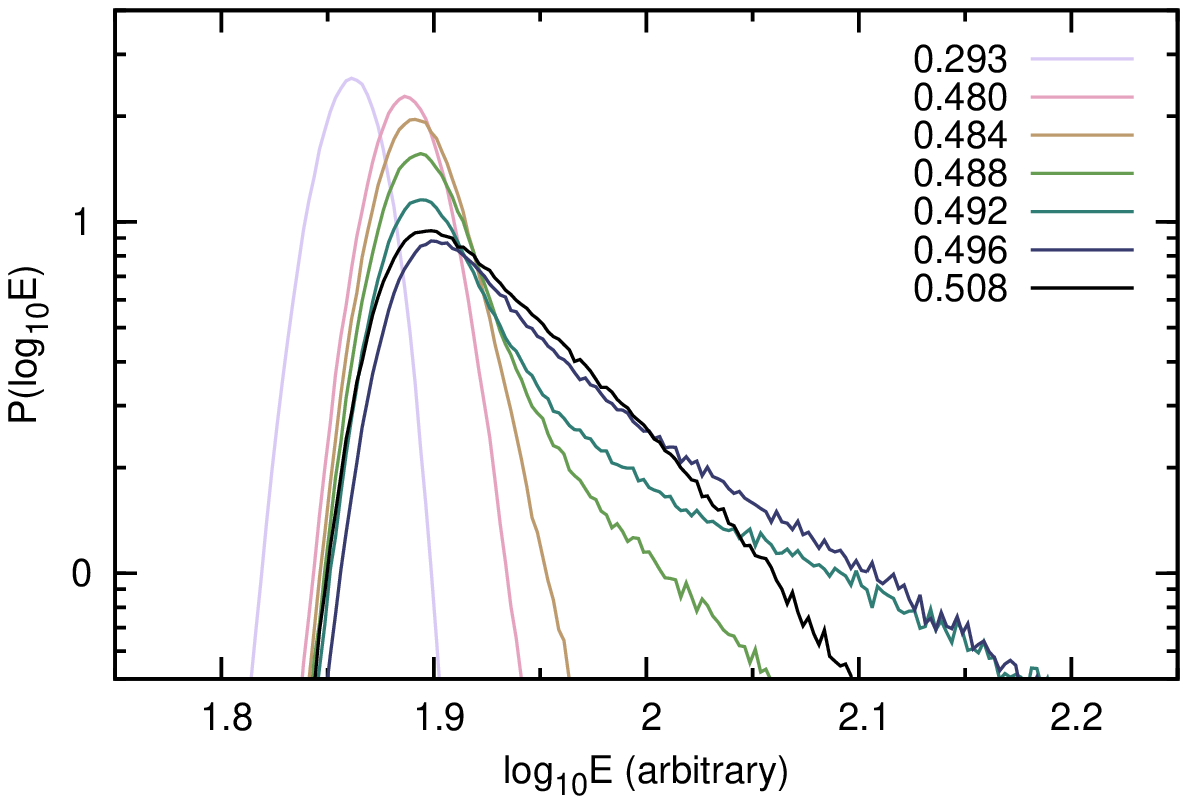}
\includegraphics[angle=0,width=\columnwidth]{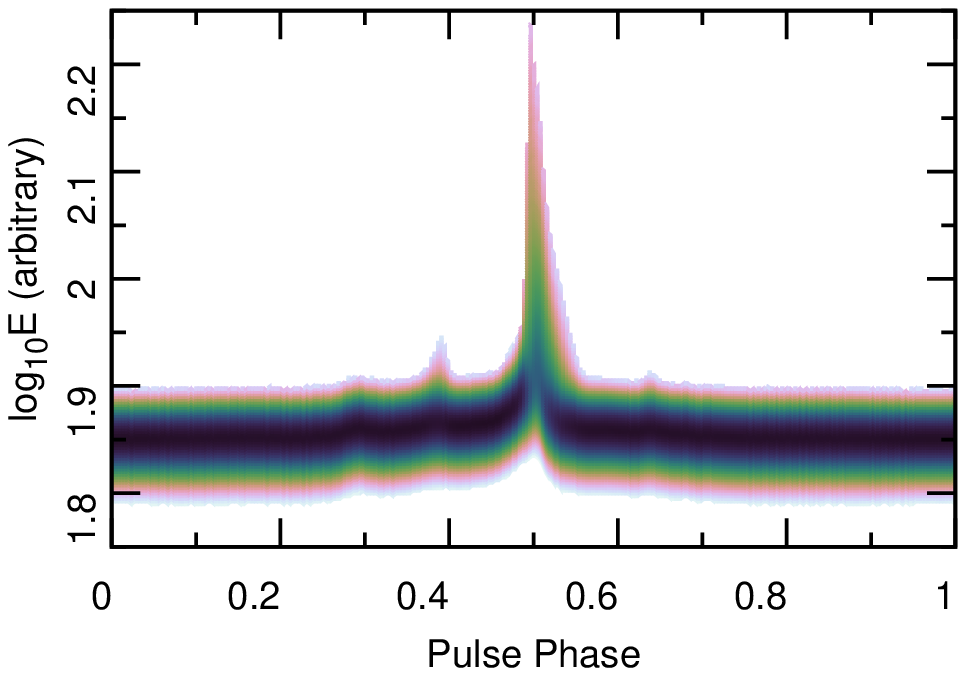}
\caption{Top: phase-resolved electric field magnitude probability density function. PDFs in seven chosen phase ranges, each integrated over $1/256$ turns of phase and centred on the values denoted in the top right of the plot, are shown. For the vast majority of pulse phases (not shown) the distribution is very similar to that of the phase range centred on 0.293. The remaining plotted phase ranges are all near the peak of the pulse profile. Bottom: the same quantity but colour coded and plotted as a function of pulse phase. All the visible excursions towards higher values correspond to peaks of the average pulse profile.\label{Fig::SGT}}
\end{figure}

While power-law distributions of the electric field have been interpreted in the past as evidence of non-linear processes in the pulsar's magnetosphere, the observed indices did not agree with theoretical predictions. \citet{2003MNRAS.343..523C} proposed an explanation for this discrepancy. They argue that power law distributions can be mimicked by two plasma wave populations that follow either log-normal statistics, as predicted by SGT, or normal statistics. \citet{2002PhRvE..66f6614C} present a convolution technique that also applies to electromagnetic fields originating from different source regions. This explains the unexpected indices of power-law distributions without the need to invoke non-linear processes.

To calculate the distribution of the electric field's magnitude we substitute the variables and ensure proper normalization:
\begin{equation}
P\left(\log_{10} E\right) = 2E^2 P\left(I\right)\,{\rm ,}
\label{eq:plasma}
\end{equation}
where I is the intensity and $E$ is the magnitude of the wave electric field. We note that in practice, to avoid rejecting very low  flux density samples, an arbitrary offset is added to the observed  flux density, following the method of \citet{2004MNRAS.353..270C}. We are interested only in the shape of the magnitude distribution and such an absolute offset does not complicate interpretation while allowing more samples to be included in the estimates of the distribution in question.

The measured phase-resolved electric field  probability density functions (PDFs) are shown in the top of Fig. \ref{Fig::SGT} for a few chosen phase ranges.  The PDFs in the off-pulse and low  flux density regions (not shown) are well fit by a Gaussian distribution. The first three  flux density distributions shown, representative of distributions for phases where the pulsar  flux density is significant but not near its peak value, are well fit by a log-normal distribution, consistent with SGT predictions.  An investigation of both the low and high  flux density tails of the distribution reveals that a better fit may be obtained by a convolution of two populations, as suggested by \citet{2003MNRAS.343..523C}. This behaviour, already demonstrated for a number of pulsars, provides further support for the hypothesis that plasma in the pulsar magnetosphere is well modelled by SGT. The four PDFs for phases $0.488$ to $0.508$ (shown in the top of Fig. \ref{Fig::SGT}) correspond to phases near the peak of the total intensity. Strong deviations from log-normality approaching a power law distribution are visible, with a significant population of pulses with very high magnitudes. We note that the  flux density distribution in the notches, i.e., the ``w'' shaped feature \citep[see e.g.,][]{2007A&A...465..981D} of the average pulse profile visible around phase $0.7$ of Fig. \ref{Fig::std}, is indistinguishable from the distribution in nearby phases.

\subsection{Intensity modulation}

Many pulsars exhibit periodic intensity modulations. Sometimes the modulation function is stationary while in other cases it drifts with respect to pulse phase. The drifting phenomenon was first discovered by \citet{1968Natur.220..231D} by visual inspection of sequences of single pulses. We have searched our data for evidence of such phenomena by calculating longitude resolved fluctuation spectra \citep{1970Natur.227..692B,1970Natur.228.1297B,1973ApJ...182..245B,1975ApJ...197..481B} and two dimensional fluctuation spectra \citep[also referred to as harmonic-resolved fluctuation spectra,][]{2001MNRAS.322..438D,2002A&A...393..733E}. An implementation of both of these spectrum calculations is available as the ``drifting\_subpulses'' application within the {\sc PSRchive} software suite.

We found no evidence of either stationary or drifting quasi-periodic intensity modulation. This places constraints on some models of the \psr\ emission mechanism. \citet{1997MNRAS.285..561G} modelled \psr\ by assuming that the wings of the profile are the result of subpulse drift in the conal component of the beam. The lack of sub-pulse drift is consistent with a previous study by \citet{1998ApJ...501..823V} at a frequency of $327$~MHz. In the model of \citet{2012MNRAS.423.3502J} lack of drifting in \psr\ implies that this pulsar spins such that the relative orientation of rotation spin and polar-cap magnetic  flux density is positive; i.e., the axes are closer to being aligned than anti-aligned; as it is the opposite direction of spin that provides the physical basis for phenomena such as nulling, drifting, and mode switching. In brief, this is related to the physics of the polar cap; no existing model can explain the existence of electron-positron pair production regions (known as sparks) in the cases of aligned magnetic and spin axes. In the opposite case, their existence and, as a consequence, the phenomenon of nulling and drifting comes naturally \citep{1975ApJ...196...51R,2010MNRAS.401..513J,2011MNRAS.414..759J}.

\begin{figure}
\includegraphics[angle=-90,width=\columnwidth]{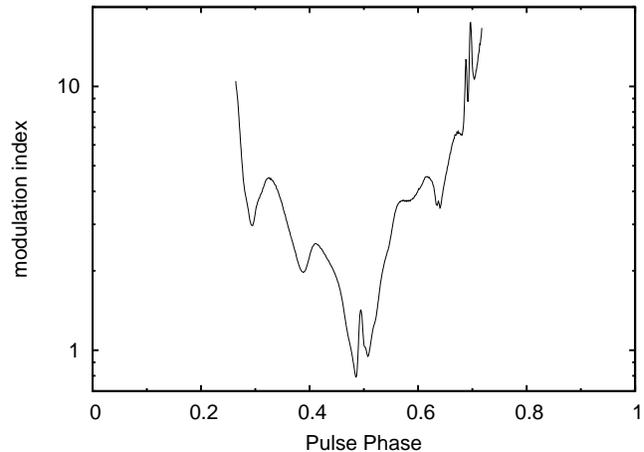}
\caption[The phase-resolved modulation index of \psr]{The phase-resolved modulation index of \psr. \label{Fig::modulation}}
\end{figure}

To study the non-periodic modulation, we present in Fig. \ref{Fig::modulation} the phase resolved modulation index, i.e., the standard deviation of the intensity normalized by the mean intensity. The on-pulse regions on the edge of the profile, where the average  flux density is low, show very high modulation. This is a typical behaviour seen in many other pulsars \citep[see e.g.,][]{1975ApJ...195..513T,1983ApJ...265..372K,2004ApJ...606.1154M}. The central region of the pulse profile shows a local maximum of the modulation. We note that all of the values are above the critical value of $0.3$ which corresponds to purely Gaussian intensity modulation \citep{2004ApJ...606.1154M}.

\begin{figure}
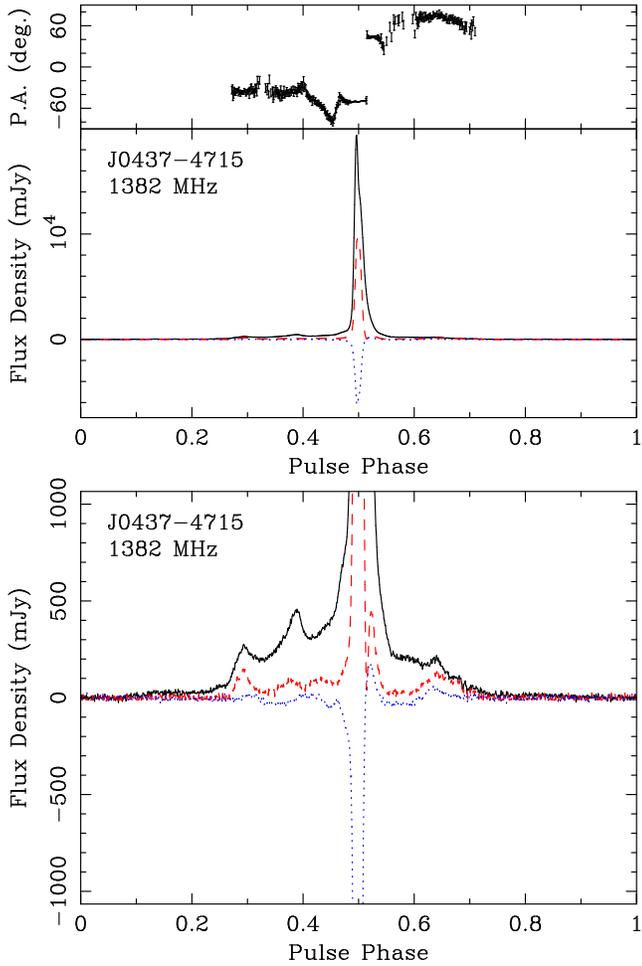
   
\includegraphics[angle=-90,width=\columnwidth]{figures/brightest_1percent}
\includegraphics[angle=-90,width=\columnwidth]{figures/brightest_1percent_crop}
\caption[The average pulse profile of the brightest one per cent subpulses]{Top: the average pulse profile of the brightest one per cent of subpulses. Bottom: zooming in on the lower parts of the profile reveals that all the Stokes parameters are consistent within the noise with the template. Meaning of various lines is the same as in Fig. \ref{Fig::std}. \label{Fig::brightest_1p}}
\end{figure}

\section{Polarimetry}
\label{pip}

Pulsars are highly polarized sources and no study of pulsar emission can be complete without a description of the polarized  flux density. In this section we analyse the properties of  the polarized emission of \psr.  We begin by discussing the dependence of polarisation on the signal-to-noise ratio. Afterwards we describe the detection of orthogonally polarised modes.

\subsection{polarization dependence on $S/N$}
\label{pol_dep_sn}
As in section~\ref{avg_SN}, we now investigate the polarization properties of the single pulses in chosen $S/N_n$ ranges. As previously noted, when average profiles are formed by selectively integrating single pulses, the structure of the wings of the profile remains similar, with only the ratio of the peak  flux density to  flux density in the wings changing. In particular, the brighter pulses have more  flux density between the local peaks of total intensity in the wings than the weaker pulses. An example of the pulse profile constructed from the 1\% brightest pulses as defined by their $S/N_{\rm peak} $ is shown in Fig.~\ref{Fig::brightest_1p}. Here, the Stokes parameters in the wings of these pulses are consistent with those of the template profile at most phases. However, the brightest pulses have a different polarisation profile and degree of polarisation in the centre of the averaged pulse profile as readily visible by comparing the bottom panel of Fig.~\ref{Fig::brightest_1p} with Fig.~\ref{Fig::std}. The low level emission appears independent of the emission in the central part as its properties are consistent with those of the template profile. The PA curve of the brightest pulses also follows the relevant curve of the template profile, except that the transition between the orthogonal modes occurs at an earlier phase in Fig. \ref{Fig::std}, i.e., later phase in Fig.~\ref{Fig::brightest_1p}.

\begin{figure}
\includegraphics[width=\columnwidth]{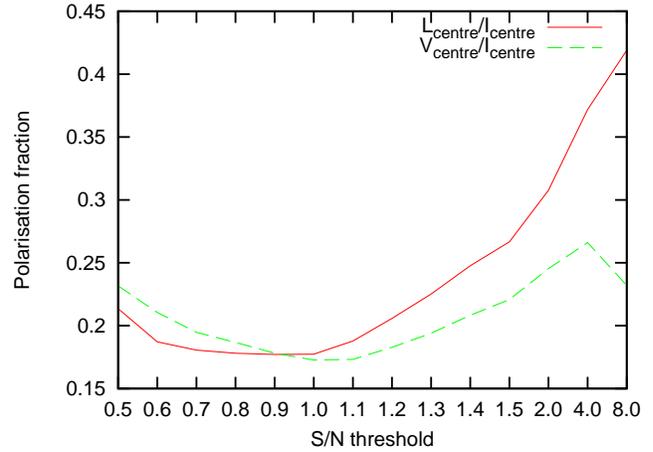}
\caption[polarization fraction as a function of $S/N_n$]{Polarisation fraction as a function of $S/N_n$. The red solid and green dashed lines correspond to the degree of linear and circular polarisation, respectively. \label{Fig::pol_degree}}
\end{figure}

The change in the polarization degree is also visible when we investigate it in the data formed from selective integration in the same $S/N_n$ regimes as considered previously. Fig.~\ref{Fig::pol_degree} shows how the degrees of linear and circular polarisation change across the integration ranges within 40 phase bins from phase $0.459$ to $0.498$, i.e. where the biggest difference in polarised pulsed profiles in Fig.~\ref{Fig::brightest_1p} is visible. Note that the degree of linear polarisation is calculated as $\left(\sum_i \sqrt{Q_i^2+U_i^2}\right)/\left(\sum_i I_i\right)$ and that of circular polarisation as $\left(\sum|V_i|\right)/\left(\sum_i I_i\right)$), where summation is over phase bins in the aforementioned phase range and no bias removal was performed. However, because we calculate the degree of polarisation based on the averaged profiles, there is no significant bias in the measurement. The steady rise of polarised  flux density in the central part of the pulse profile with increasing $S/N_n$ threshold is apparent, possibly due to the enhanced coherence of the emission mechanism responsible for the brighter pulses. 

\begin{figure*}

\includegraphics[angle=0,width=0.49\textwidth]{figures/pulse_example_S_TL.ps}
\includegraphics[angle=0,width=0.49\textwidth]{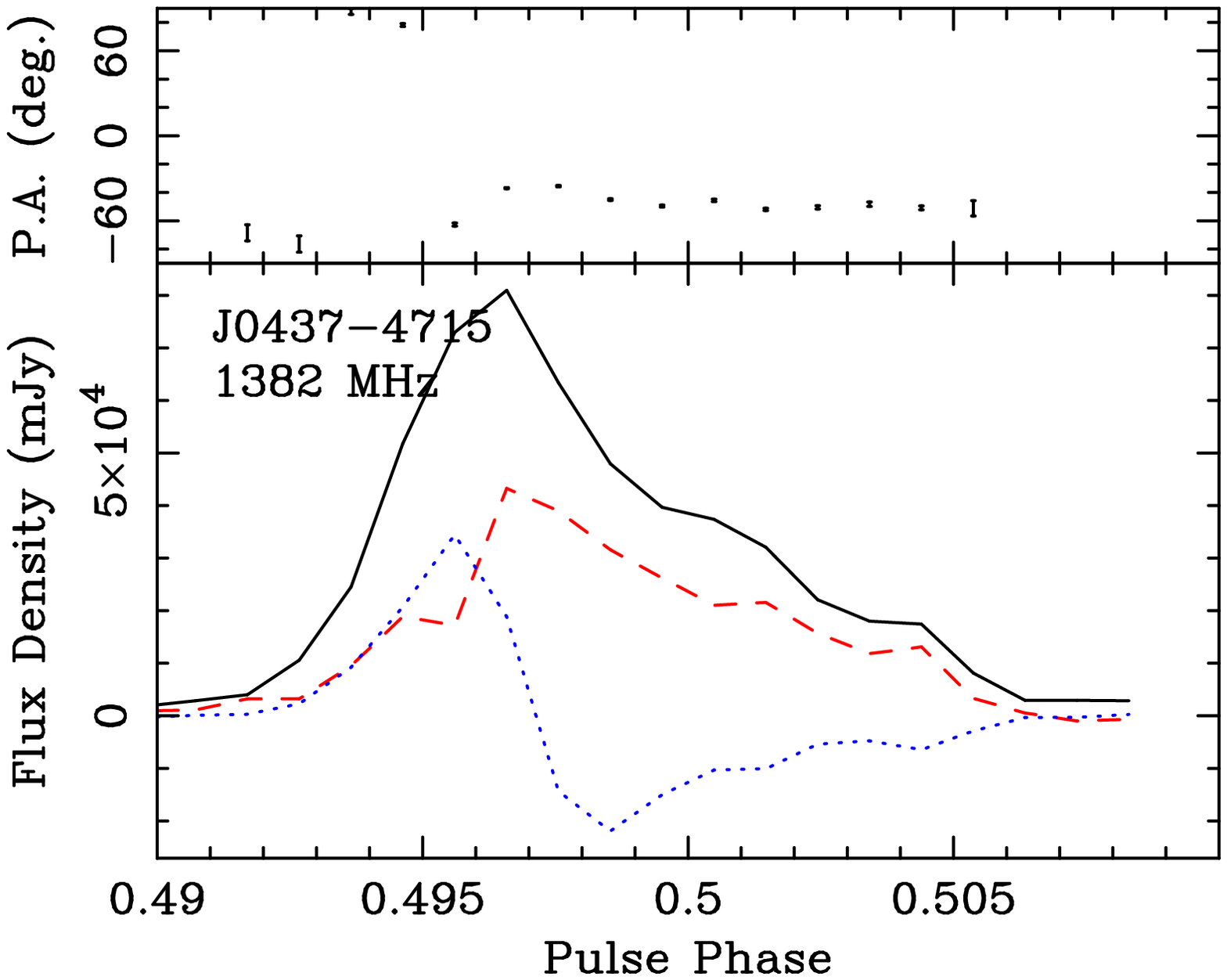}
\includegraphics[angle=0,width=0.49\textwidth]{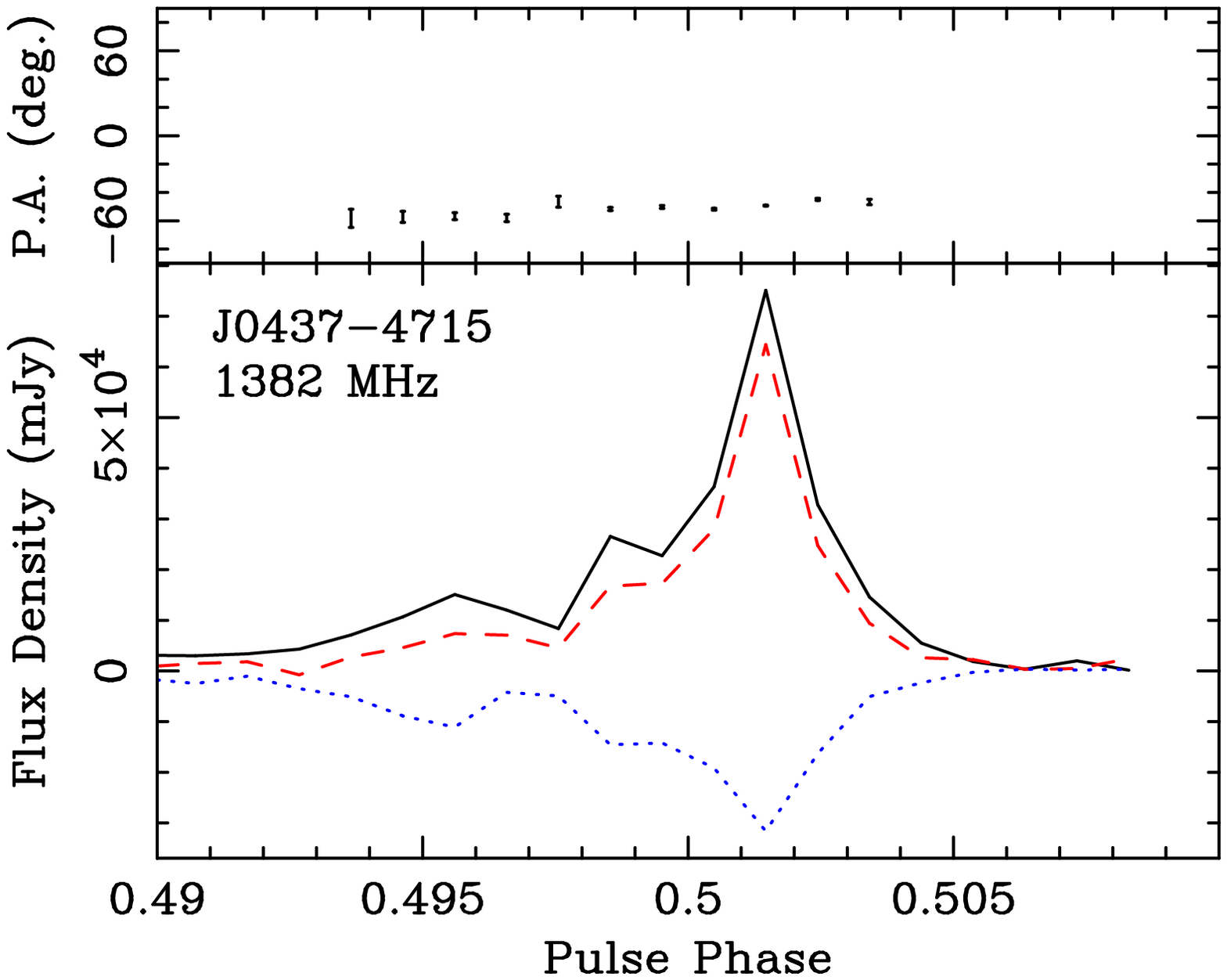}
\includegraphics[angle=0,width=0.49\textwidth]{figures/pulse_example_S_BR.ps}
\caption[Examples of bright subpulses]{Examples of bright subpulses. The two pulses in the top exhibit the swing of circular polarization. See the text for more details. \label{Fig::examples}}
\end{figure*}

\citet{2009ApJ...696L.141M} analysed the polarization of single pulses from 10 slowly rotating pulsars. They found that all of the objects in their sample emit bright pulses that have a high degree of linear polarisation and a change of the sense of circular polarisation across the sub-pulse. These features may originate in coherent curvature radiation from a single charged particle travelling at relativistic speed in the presence of a magnetic field \citep{1987ApJ...322..822M}, producing an escaping extraordinary plasma wave \citep{1986ApJ...302..120A}. \citet{2009ApJ...696L.141M} argue that a soliton is a good candidate for such a particle. A sample of bright pulses, as selected from the top one per cent of the brightest pulses according to their $S/N_{\rm peak}$, is shown in Fig.~\ref{Fig::examples}. The two top pulses exhibit the same characteristics that \citet{2009ApJ...696L.141M} discussed. While difficult to estimate reliably, the fraction of pulses with such a feature appears to be very small. Among the brightest one per cent of pulses only about half of them exhibit it. The fraction of such pulses appears to diminish among weaker pulses but is increasingly difficult to estimate robustly. The example in the bottom left panel shows a pulse that has a high degree of linear polarisation but Stokes V does not change sense across the pulse; the bottom right panel is an example of a sub-pulse with a lower degree of polarisation. The data indicate that at least some of \psr\ emission can originate from coherent curvature radiation of solitons. We stress that only the pulses occurring just before the phase 0.5 exhibit a swing of circular polarisation, that is the emission which possibly originates from solitons is detected in a limited range of pulse phase.

\begin{figure*}
\includegraphics[width=\textwidth]{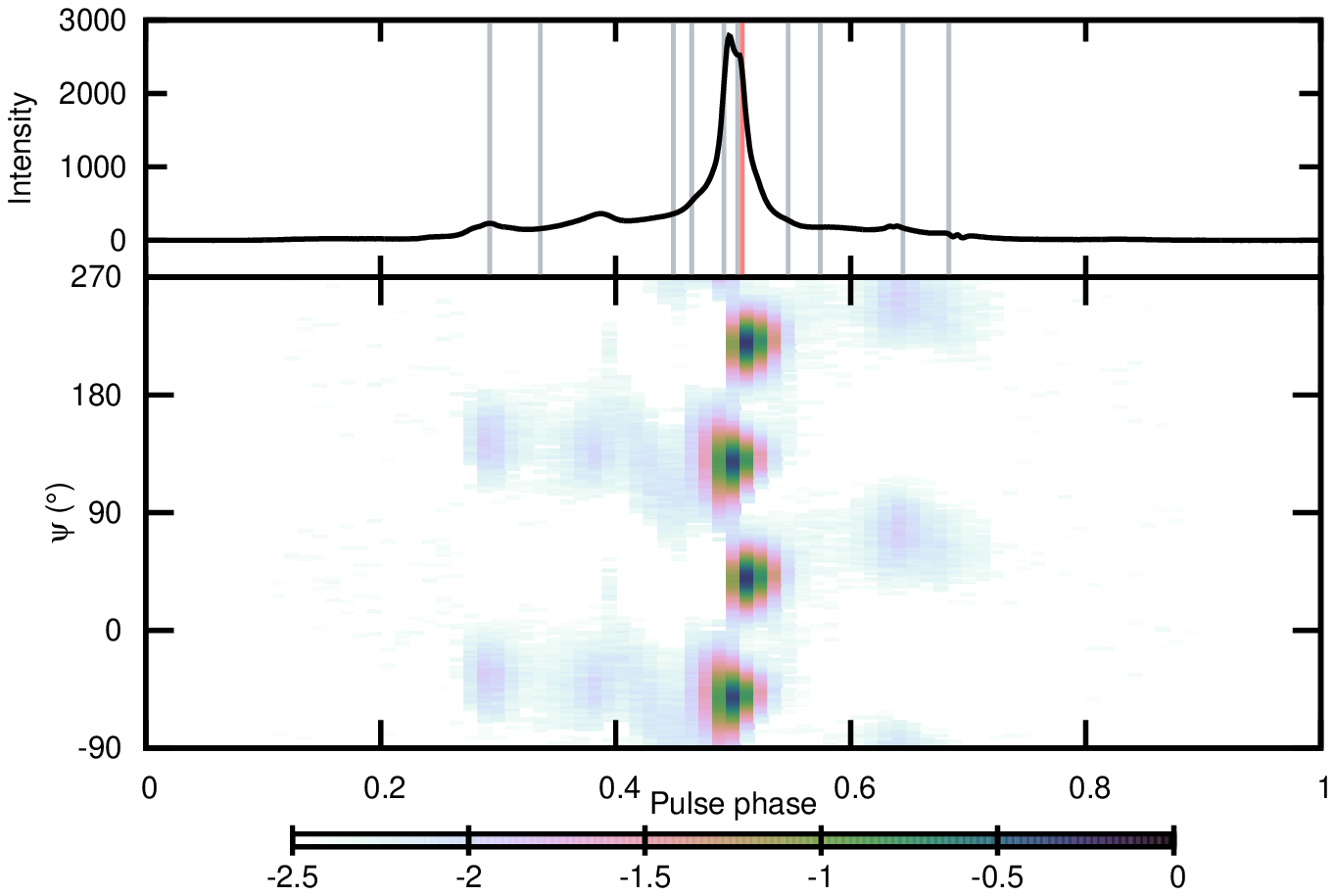}
\caption[Phase resolved distribution of polarization angle.]{Logarithm of the phase resolved distribution of polarisation angle is shown in the bottom panel while the top panel shows a total intensity template with a few chosen phases of interest marked in grey and red.\label{Fig::PA_ma}}
\end{figure*}

\subsection{Phase resolved polarization histograms}

We also analysed the polarimetric properties of \psr\ while preserving phase resolution. In this section we present figures similar to Fig. 1 and 2 in \citet{2004A&A...421..681E}. We first present the distribution of polarization angles derived from every single pulse in Fig. \ref{Fig::PA_ma}. The top panel shows the total intensity template profile with a few special phases marked with grey and red  highlights. The bottom plot shows an unweighted phase-resolved histogram of polarisation angles, normalized by the maximum value across all phases, with the perceived intensity of the image linearly dependent on the value of the histogram bin in both colour and black-and-white print, using the Cubehelix colour coding scheme \citep{2011BASI...39..289G}. The polarisation angle is plotted twice for clarity. We point out the unusual distribution of the polarisation angle near the peak of the pulse profile and polarisation angle values close to $45$ degrees.

\begin{figure*}
\includegraphics[width=\textwidth]{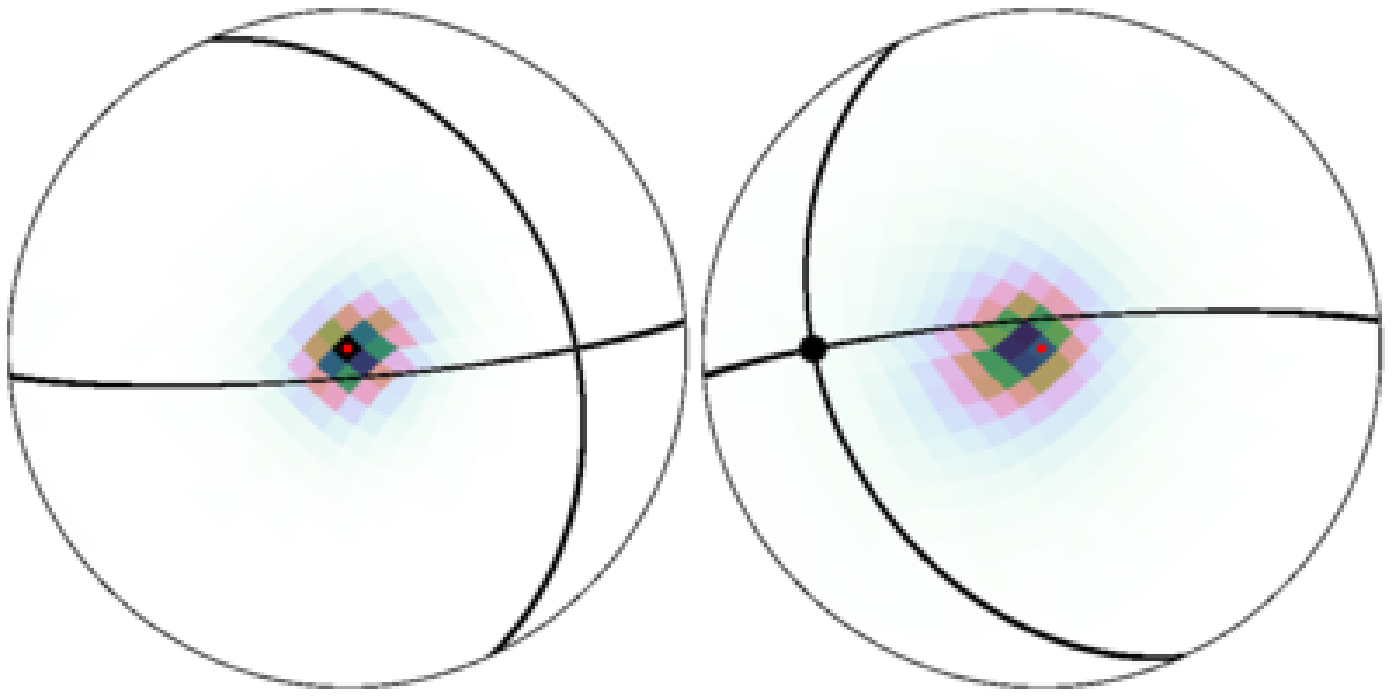}
\caption[Lambert projection of the Poincar{\'e} sphere - OPMs]{Lambert equal area azimuthal projection of the Poincar{\'e} sphere interrupted at the equator at phase $0.508$, marked in red in Fig. \ref{Fig::PA_ma}. To avoid heavy distortions, the projection is interrupted at the equator and the other hemisphere is plotted on the right-hand side. The intensity of the sphere is proportional to the value of the histogram in the given point on the Poincar{\'e}'s sphere. Orthogonally polarized modes are clearly visible. The black lines and the large black dot at their intersection correspond to four meridians (longitudes of 0, 90, 180, and 270 degrees) and the north pole of the Poincar{\'e} sphere, which corresponds to fully left hand circularly polarised radiation. The red squares mark the poles of each hemisphere. Colour coding is the same as in Fig. \ref{Fig::PA_ma}. \label{Fig::opm_orth}}
\end{figure*}

\begin{figure*}
\includegraphics[width=\columnwidth]{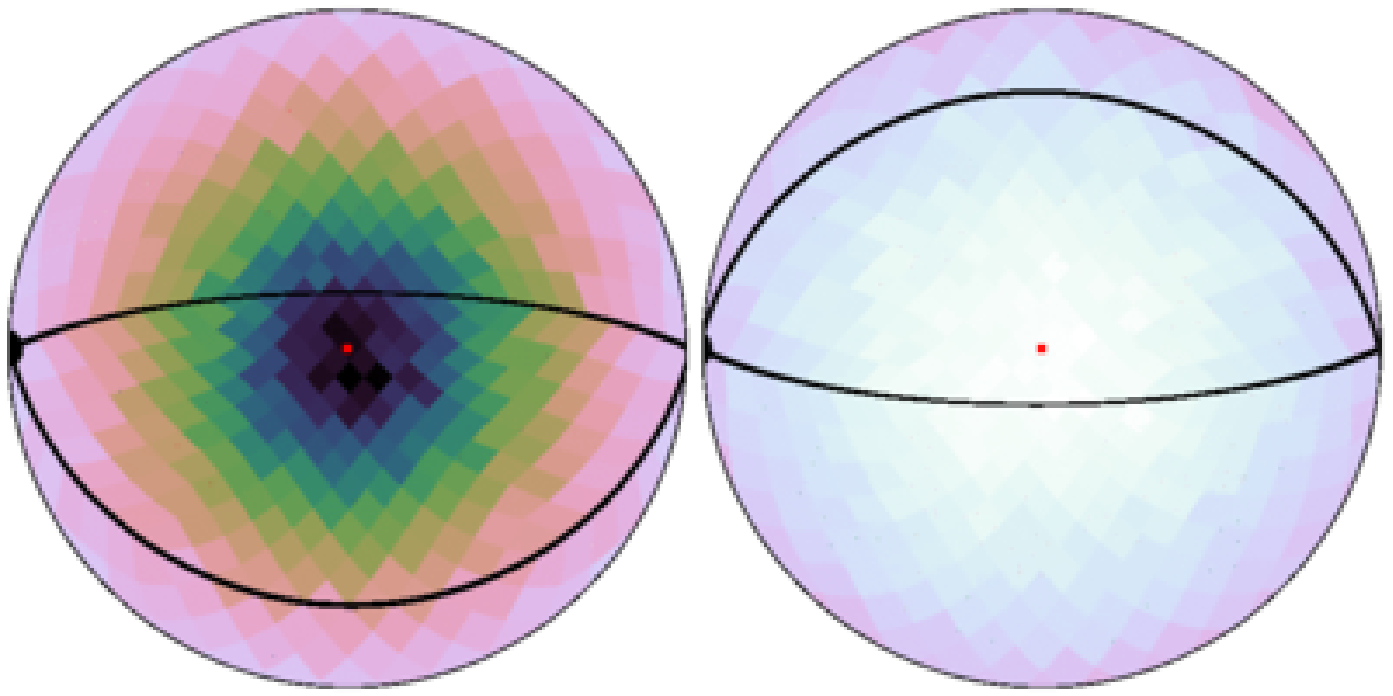}
\includegraphics[width=\columnwidth]{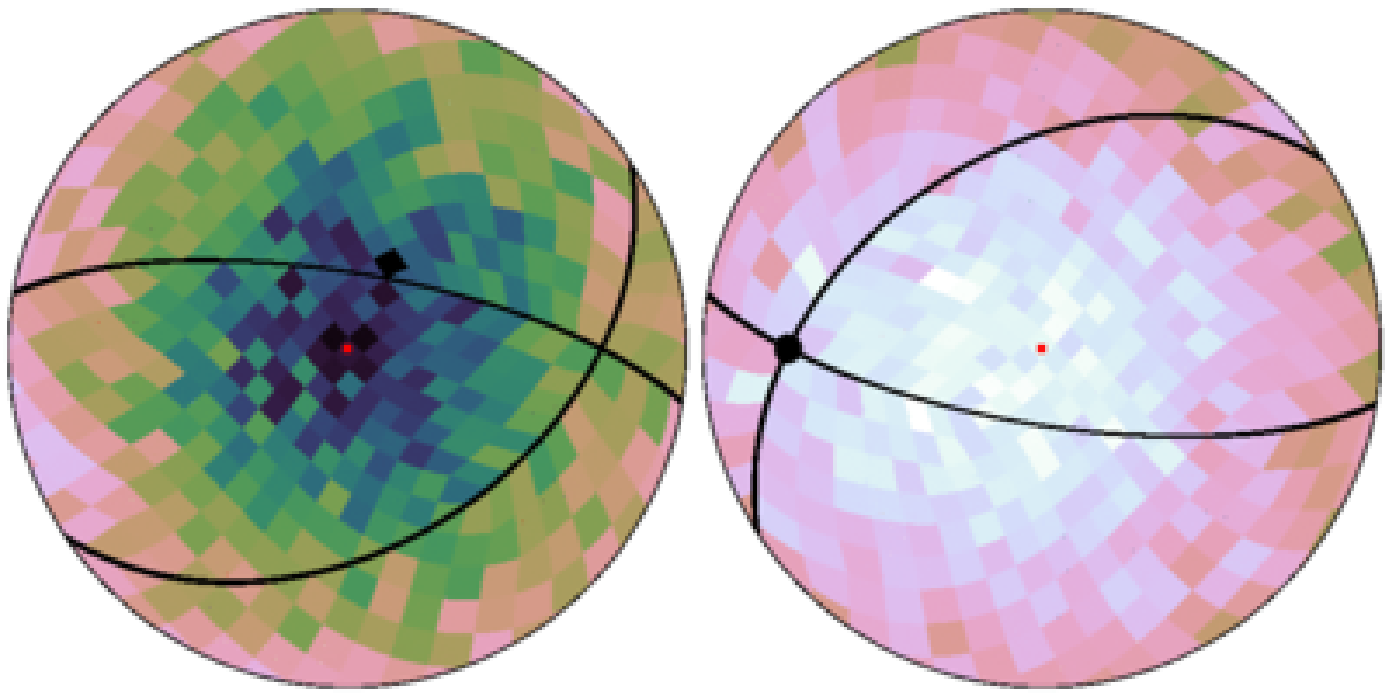}
\includegraphics[width=\columnwidth]{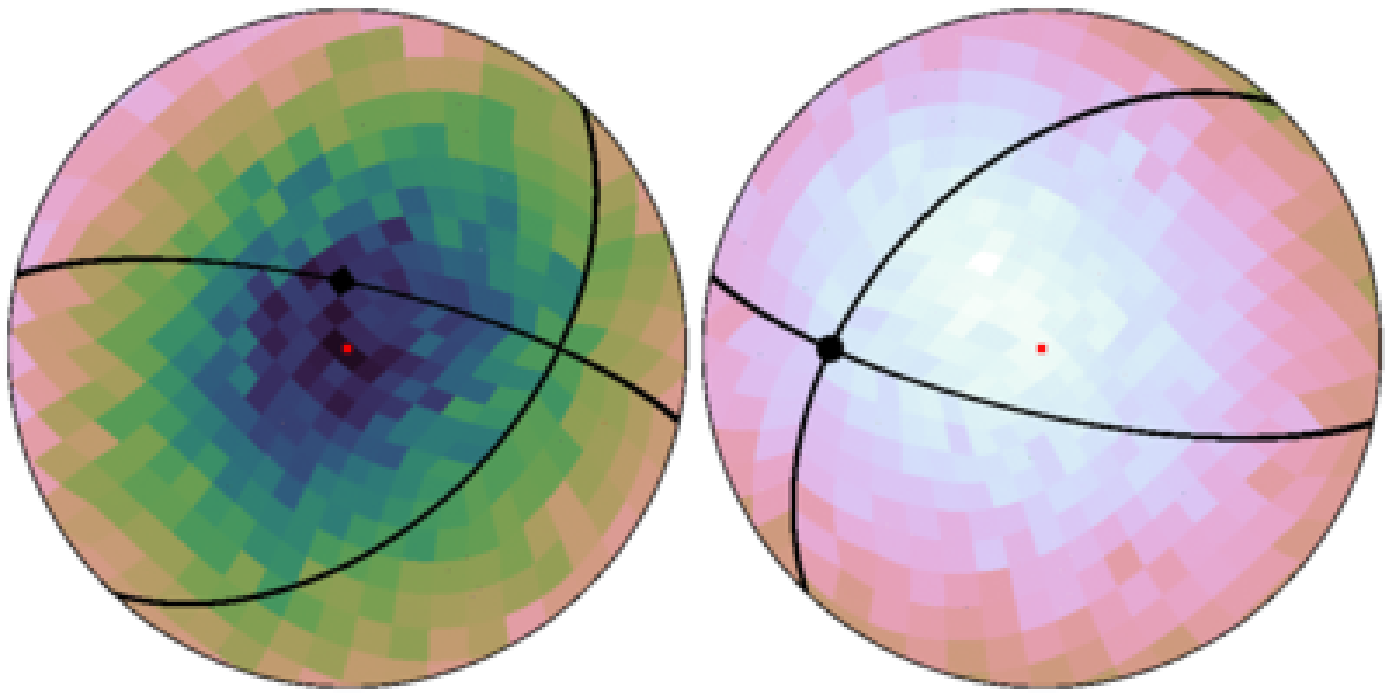}
\includegraphics[width=\columnwidth]{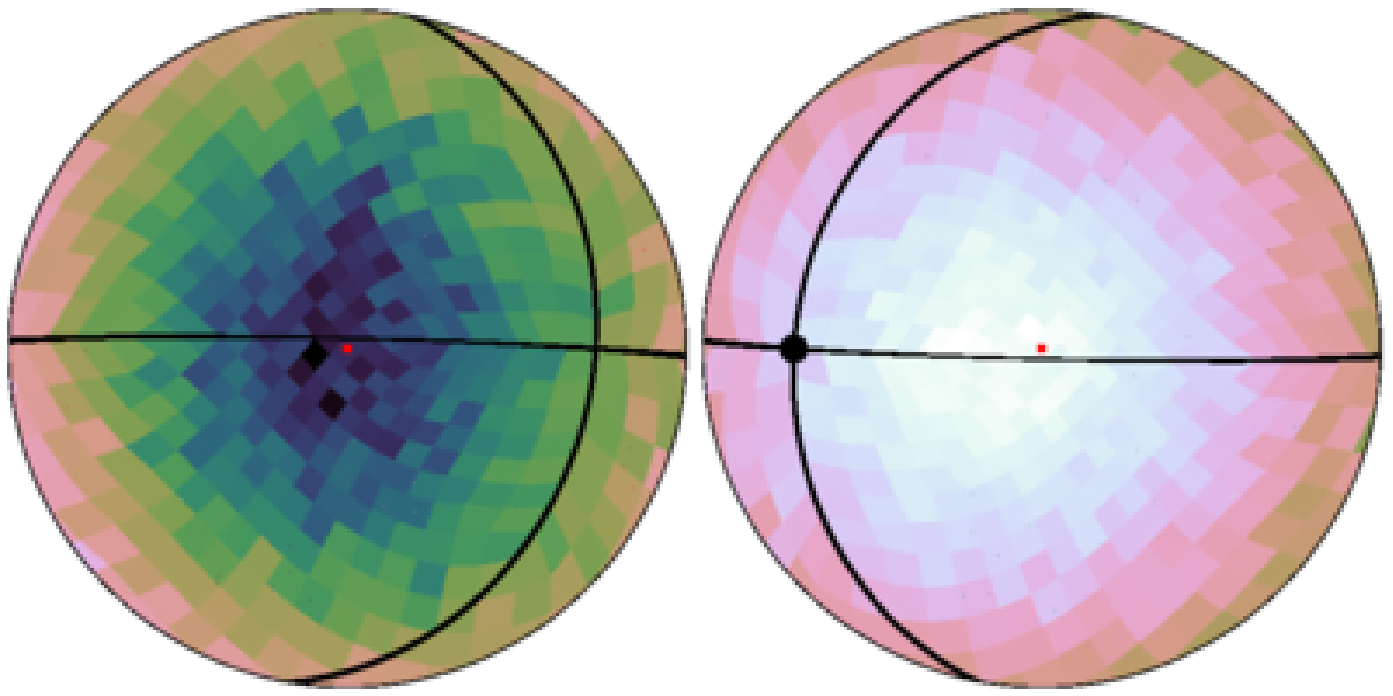}
\includegraphics[width=\columnwidth]{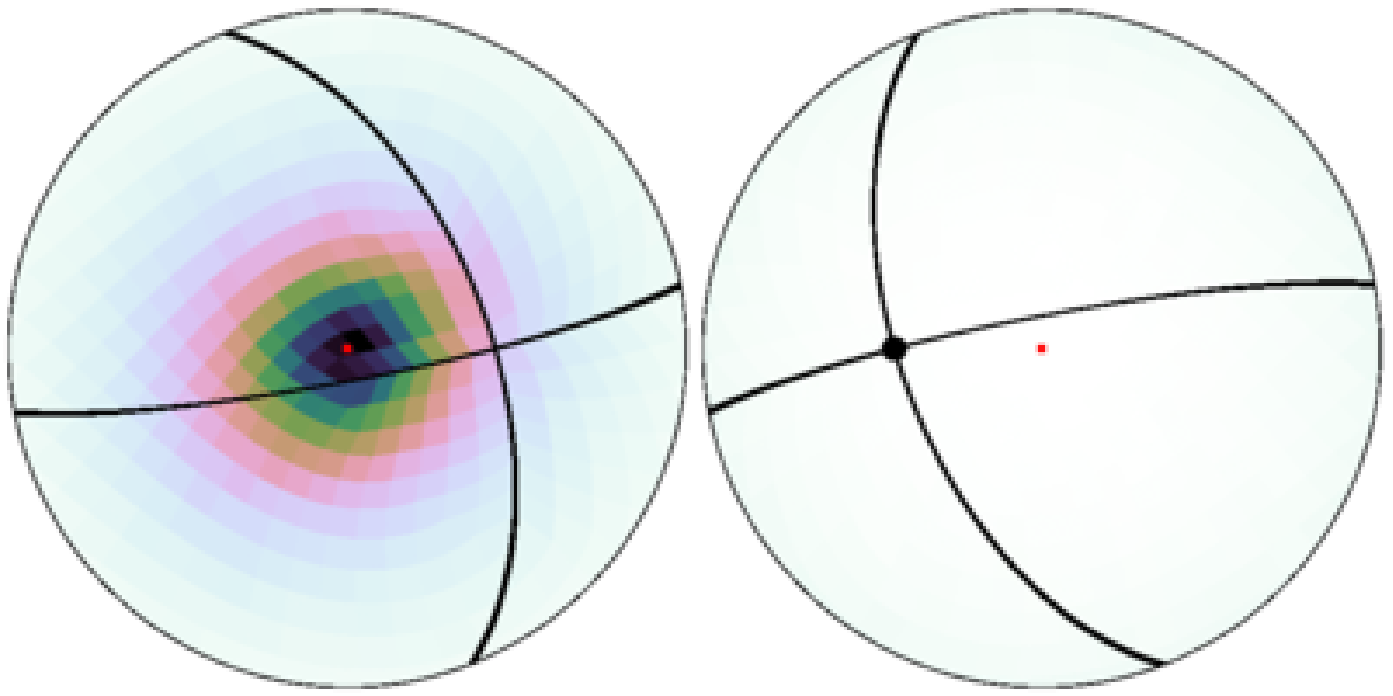}
\includegraphics[width=\columnwidth]{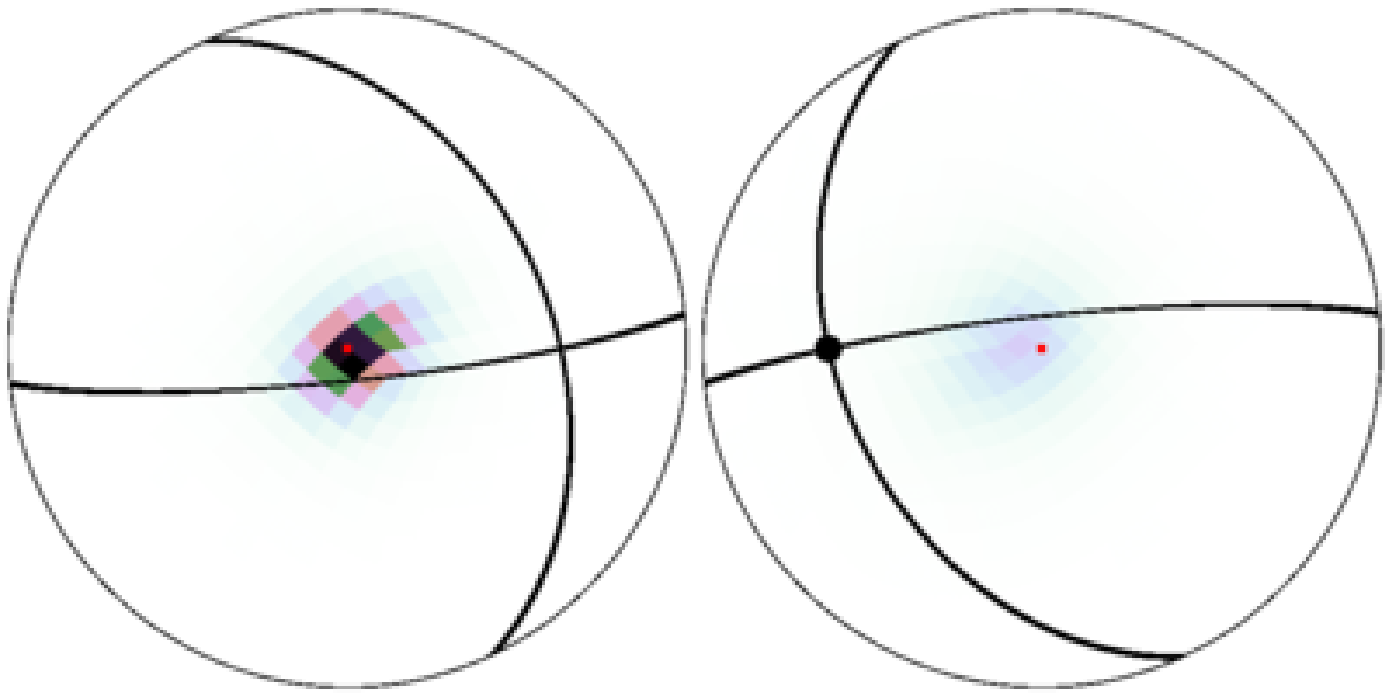}
\includegraphics[width=\columnwidth]{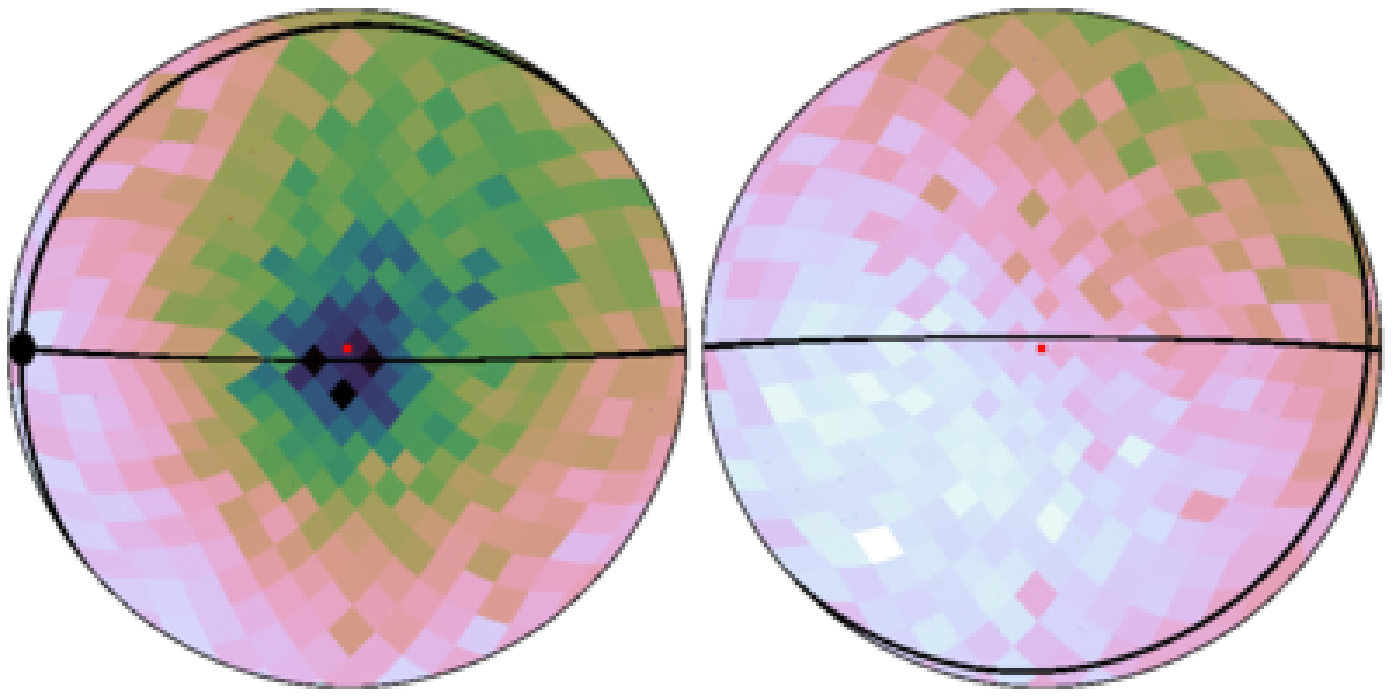}
\includegraphics[width=\columnwidth]{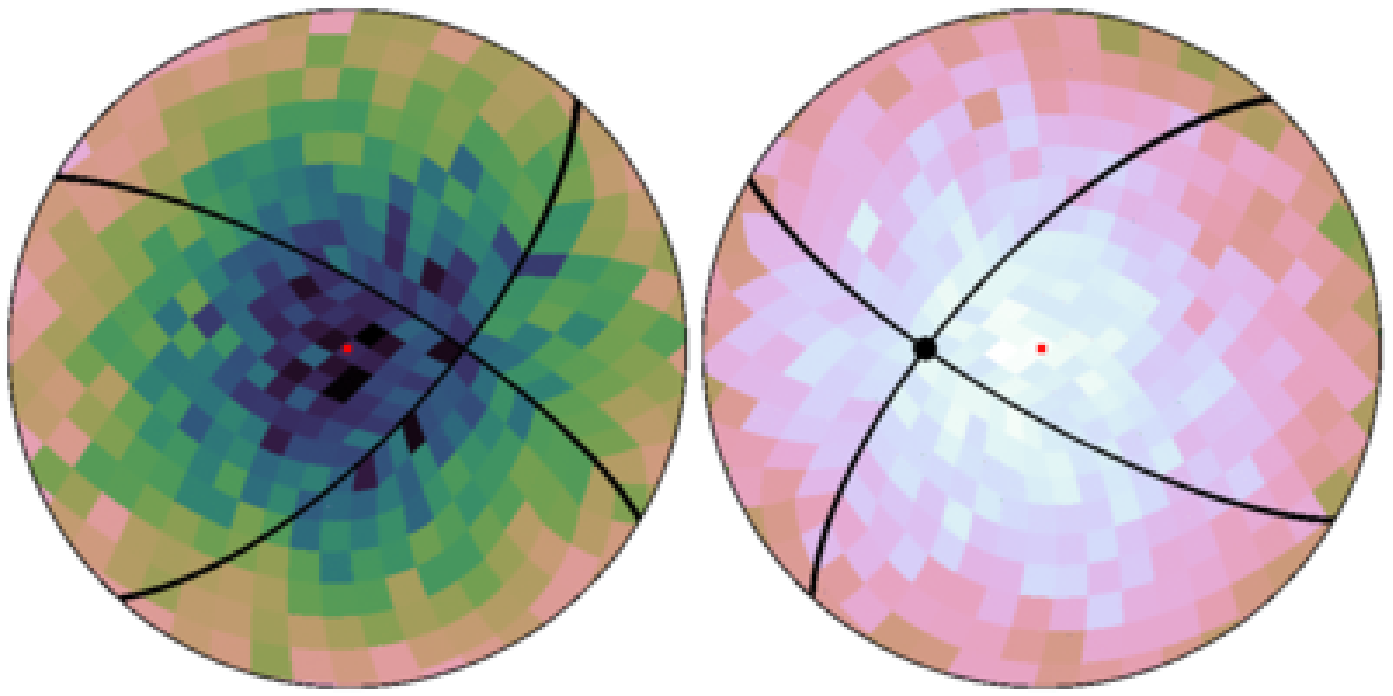}
\includegraphics[width=\columnwidth]{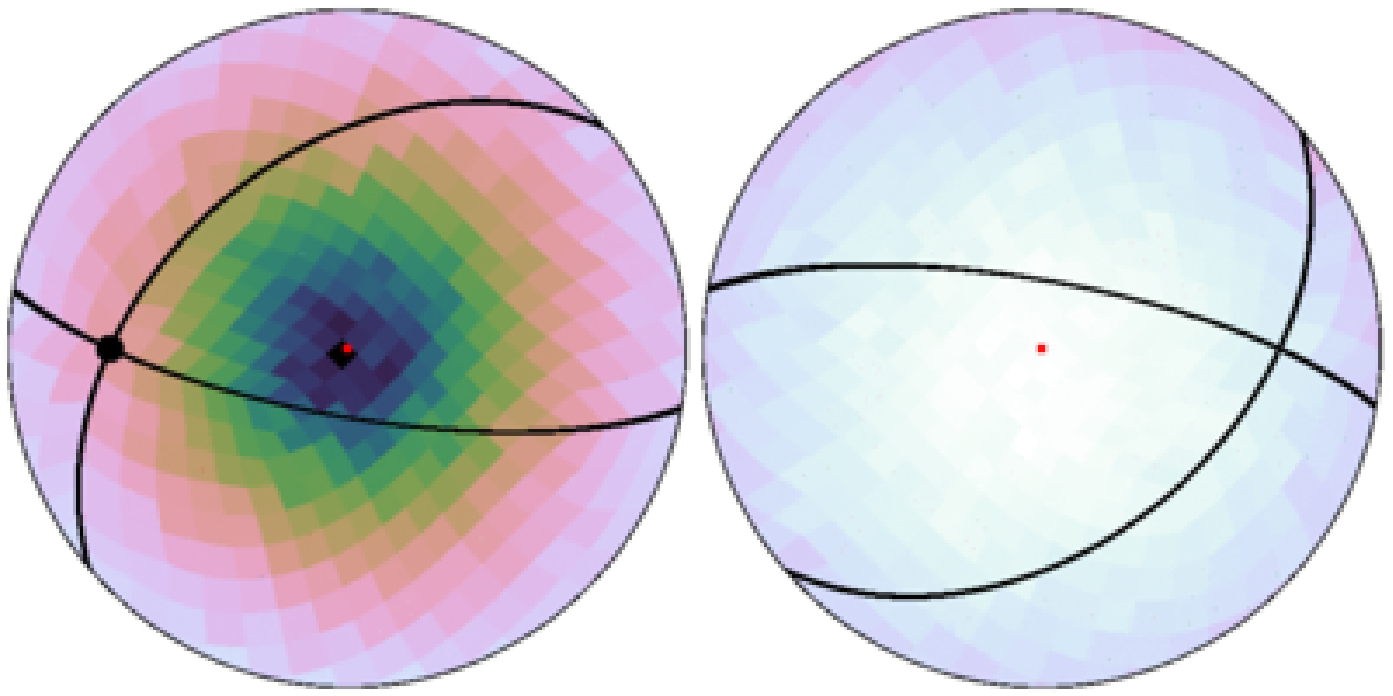}
\includegraphics[width=\columnwidth]{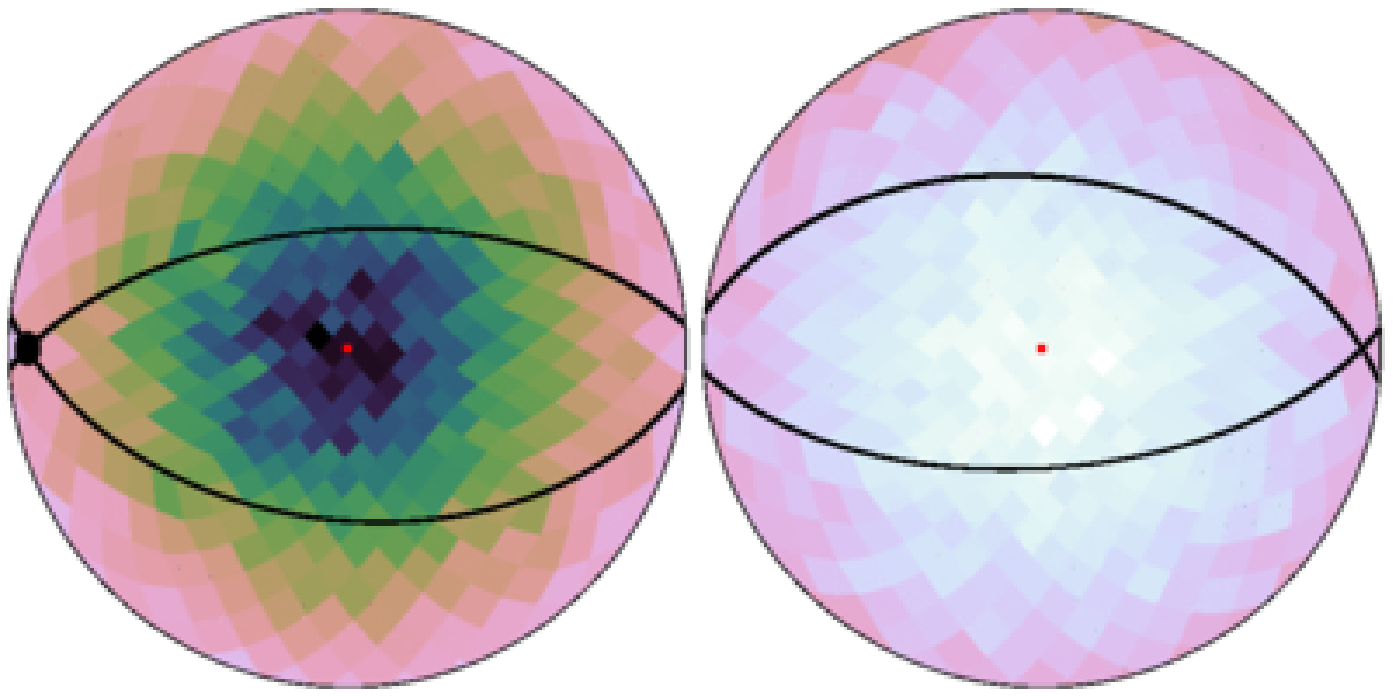}
\caption[Lambert projection of the Poincar{\'e} sphere, part 1]{Lambert equal area azimuthal projection of the Poincar{\'e} sphere interrupted at the equator near the phases 0.293, 0.336, 0.449, 0.465, 0.492, 0.504, 0.547, 0.574, 0.645, 0.684, marked in Fig. \ref{Fig::PA_ma} in gray. The projections are presented in reading order with two projected hemispheres per phase range. Colour coding and meaning of all marks is the same as in Fig. \ref{Fig::opm_orth}.
\label{Fig::opm_pl}}
\end{figure*}

We also calculate the phase-resolved histograms of Stokes parameters on the Poincar\'{e} sphere. We implemented this using the Hierarchical Equal Area isoLatitude Pixelation of a sphere \citep[HEALPix,][]{2005ApJ...622..759G} with the resolution parameter set to three, corresponding to 768 pixels on the sphere. The generation of Poincar\'{e} sphere histograms was performed with the ``psrpol'' application within the {\sc PSRchive} framework. This application reads in every subpulse and, for each pulse phase, computes the appropriate pixel on the sphere corresponding to the polarization vector and weights the contribution to that pixel by the polarised  flux density. In this way we obtain a separate histogram of polarisation vector for all pulse phases. We present a few chosen Poincar\'{e} spheres in Fig. \ref{Fig::opm_orth} and \ref{Fig::opm_pl} (an animation showing the evolution of the Poincar\'{e} sphere with the rotation phase can be obtained from the corresponding author). The first of these figures corresponds to the phase marked in red in Fig. \ref{Fig::PA_ma}, while the second one shows the remaining phases marked in Fig. \ref{Fig::PA_ma} in grey. We note that the mean of the polarisation vector varies significantly across the pulse phase, well beyond the variance of the polarisation at any phase.

The most notable feature of the polarization patterns visible in Fig. \ref{Fig::PA_ma}, \ref{Fig::opm_orth} and \ref{Fig::opm_pl} is the presence of orthogonally polarised modes \citep{1984ApJS...55..247S} near the peak  flux density of the total intensity at the phase $0.508$. The modes appear to be not perfectly orthogonal.  While the presence of OPMs is expected for this pulsar based on its average profile \citep{1997ApJ...486.1019N}, to our knowledge this is the first direct detection of OPMs in a millisecond pulsar. More surprisingly, we do not detect OPMs at different phases, where the emission is dominated by one mode despite the complicated shape of the polarisation angle curve across spin phase. This suggests that the emission may originate in several regions \citep{1997ApJ...486.1019N}, in contrast to the conclusion of \citet{1997MNRAS.285..561G}. Furthermore, based on the presence of OPM near the peak of the pulse profile and the rapid swing of polarisation angle shown in Fig. \ref{Fig::PA_ma}, we support the idea that the impact parameter is of the order of a few degrees \citep{1997ApJ...486.1019N}, consistent with the fit of a relativistic rotating vector model to the polarisation angle sweep \citep{1997MNRAS.285..561G}. In such a case, the variations of polarisation angle across the spin phase can be caused by the emission originating close to the surface, where deviations from dipolar structure of the magnetic field is still strong, or from deformations further out in the magnetosphere \citep{1991ApJ...366..261R}.

We also do not see annuli  as seen by \citet{2004A&A...421..681E} in the case of \mbox{PSR B0329+54}; however we do see deviations from a Gaussian distribution of polarization, for example in the fourth-last panel in Fig. \ref{Fig::opm_pl}, corresponding to the spin phase equal $0.547$. Detection of the modes near the peak intensity is expected from the shape of the polarisation angle curve and the increased modulation index.  The association of OPMs with highly modulated emission has been first noted by \citet{2004ApJ...606.1154M} and explained statistically by \citet{2009ApJ...694.1413V}. Furthermore, \citet{2007A&A...465..981D,2010MNRAS.401.1781D} argue that \psr\ is likely to exhibit OPMs and that the trailing side of the pulse profile  is dominated by the extraordinary polarisation mode, as this mode can explain the creation of the notches by assuming curvature emission from thin plasma streams. We note that soliton emission seen only in the leading part of the profile is also expected to result in the extraordinary mode.

We can now tie together several of our results, namely that the brightest pulses preferentially occur in the leading component, switch to a different polarization mode at a later phase, and are more polarised. By selectively integrating the brightest pulses we extend the phase region in which their polarisation properties dominate the average profile, thus delaying the transition to the second polarisation mode. Finally, since the brightest pulses occur preferentially where one of the modes dominates, their average profile exhibits less depolarisation by the superposition of OPMs.

\section{Conclusions}
\label{conclusions}

We have presented a study of \psr\ subpulse properties in the 21-cm band. We attempted to reduce the post-fit arrival time residual by minimising SWIMS via rejection of the subpulses with the highest flux density while taking scintillation into account. While we have not achieved a significant improvement in the timing precision, we demonstrated that the rms of the timing residual and the statistical goodness of the timing model fit can be improved by rejecting the brightest subpulses and demonstrated a correlation between the average ToA and the $S/N$ of the subpulses as a consequence of the leading component dominating the mean profile of brightest pulses. Phase resolved histograms of the pulse intensity have log-normal distributions in a range of pulse phase, consistent with predictions of the stochastic growth theory of plasma waves. We discovered that there is a strong increase of the degree of polarization, both linear and circular, with the pulse intensity. A special class of the brightest pulses has polarisation features compatible with those expected from a relativistic soliton. Finally, we presented the phase-resolved histograms of polarisation angle and polarisation vector. These reveal the existence of two orthogonally polarised modes in a narrow phase range, for the first time directly detected in a millisecond pulsar. The modes show slight deviation from orthogonality. Throughout most of the pulse phase, the emission is dominated by one mode and the other is not visible, if present.

\section*{Acknowledgements}
We express our gratitude to the referee, Aris Karastergiou, for his useful comments on the manuscript. The Parkes Observatory is part of the Australia Telescope National Facility which is funded by the Commonwealth of Australia for operation as a National Facility managed by CSIRO. We thank the staff at Parkes Observatory for technical assistance during observations. This work is supported by Australian Research Council grant \# DP0985272. GBH is supported by an Australian Research Council QEII Fellowship (project \# DP0878388).

%bibliography:
\newcommand{\apj}{ApJ}
\newcommand{\aj}{AJ}
\newcommand{\apjs}{ApJS}
\newcommand{\apjl}{ApJ Lett}
\newcommand{\nat}{Nature}
\newcommand{\aap}{A\&A}
\newcommand{\araa}{ARAA}
\newcommand{\pra}{Phys.~Rev.~Lett.}
\newcommand{\prc}{Phys.~Rev.~C}
\newcommand{\prd}{Phys.~Rev.~D}
\newcommand{\pre}{Phys.~Rev.~E}
\newcommand{\physrev}{Phys. Rev.}
\newcommand{\mnras}{MNRAS}
\newcommand{\nar}{New Astronomy Reviews}
\newcommand{\pasp}{PASP}
\newcommand{\pasj}{PASJ}
\newcommand{\apss}{ApSS}
\newcommand{\aapr}{AAPR}
\newcommand{\aaps}{A\&AS}
\newcommand{\physrep}{Phys. Rep.}
\newcommand{\sovast}{Soviet Astron.}
\newcommand{\pasa}{Publ. Astron. Soc. Aust.}

\newcommand{\jgr}{J. Geophys. Res.}
\newcommand{\solphys}{Sol.~Phys.}

\bibliographystyle{mn2e}
\bibliography{single_pulses}

\label{lastpage}

\end{document}